\documentclass[a4paper,11pt]{article}
\usepackage{jinstpub} 
\usepackage{lineno}
\usepackage{caption}
\usepackage{subcaption}
\usepackage[subpreambles=true]{standalone}
\usepackage{tikz}
\usepackage[siunitx]{circuitikz}
\usepackage{comment}
\standaloneconfig{mode=tex}
\usetikzlibrary{arrows.meta,calc,positioning,decorations.pathmorphing,shadows}

\newcommand{\DV}[1]{\ensuremath{\Delta \mathrm{V}_{\mathrm{#1}}}}

\title{\boldmath Signal formation and induction-gap optimization in a THGEM coupled to a resistive plate anode}

\author[a,1]{A. Maity\note{Corresponding author.},}
\author[a]{L. Moleri,}
\author[a]{M. Borysova,}
\author[a]{and S. Bressler}

\affiliation{Department of Particle Physics and Astrophysics, Weizmann Institute of Science,
\\
Rehovot, 7610001, Israel}

\emailAdd{arpan.maity@weizmann.ac.il}

\abstract{
A Thick Gaseous Electron Multiplier (THGEM) detector coupled to a resistive plate anode was investigated for the first time, using an Ar:CO$_2$ (93:7) gas mixture. The resistive anode concept enabled stable operation with induction gaps significantly smaller than those typically employed in THGEM detectors, opening the possibility of improving signal formation and timing performance. The effect of the induction gap on the detector current- and charge-signal characteristics was systematically investigated. An optimal induction gap of 0.2~mm was identified based on several key signal parameters. Subsequently, detailed studies were carried out for this optimal configuration under different electric-field settings.
The detector achieved a highest peak amplitude of $\sim$56~$\mu$A with a rise time of $\sim$5~ns at \DV{THGEM}=1900~V, \DV{induction}=100~V, and a drift field of 1~kV/cm when irradiated with 5.9~keV X-rays. Under the same operating conditions, a time resolution of $\sim$6.0~ns was measured detecting cosmic muons. This study establishes a new THGEM detector configuration and provides guidance for its potential application in particle-detection systems, such as muon spectrometers and sampling elements of digital hadronic calorimeters.
}


\begin{document}
\maketitle
\flushbottom
\section{Introduction}
\label{sec:intro}

Gaseous radiation detectors are extensively used in particle collider experiments and in various civil applications, especially those requiring cost-effective coverage of large areas. Current studies of gaseous detectors -- wire chambers, Resistive Plate Chambers (RPCs), and Micro Pattern Gaseous Detectors (MPGDs) -- aim to meet increasingly challenging experimental requirements, in particular, good temporal and spatial resolution at high particle rates, together with resilience to aging in harsh radiation environments.

Many modern gaseous detectors include electrodes made of non-metallic resistive materials, which reduce the rate of electrical discharges and quench their energy, thereby protecting delicate electrodes and sensitive readout electronics. The resistive electrodes can be made of thin films deposited on thicker insulators (see, for example,~\cite{MMreview21, BENCIVENNI23, DLC_China26,Nagai:1996mf} and references therein) or plates made of electrostatic dissipative materials (see, for example,~\cite{ResGasDet18, jash2024discharge} and references therein). In the former case, the charge reaching the resistive layer during detector operation is restored through a surface current, whereas in the latter case, the current flows through the bulk of the resistive plate. Depending on the resistivity of the material and the way it is grounded, different solutions can provide stable and reliable operation, although at the expense of some efficiency loss at high radiation flux.

In a Thick Gaseous Electron Multiplier (THGEM)~\cite{bressler2023thick} detector setup, the THGEM is preceded by a conversion drift gap. Ionization electrons produced by incident radiation drift toward the THGEM and enter its holes, where charge multiplication takes place under the influence of a strong electric field. The avalanche electrons generated in this process are then extracted into a few-mm-wide induction gap and drift toward the readout anode, where they induce the electrical signal.

The subject of this work is the operation of a THGEM coupled to a resistive plate anode to study the effect of the induction gap on the avalanche gain and signal formation. The amount of electron avalanche multiplication occurring in the holes and in the induction gap is tuned by changing the electric field strength in the two regions. The time development of the induced signal on the anode is modified by changing the induction gap size. The different geometries and operation modes provide the flexibility to balance charge gain and time response of the multiplier. This allows optimization of the detector performance, given that fast rise time and increased peak amplitude lead to improved timing precision~\cite{Ferrero2021UFSD}.

The induced signals were studied using two different readout schemes: a charge-sensitive preamplifier to evaluate the integrated charge, and a broadband current amplifier to characterize the fast electron-induced current pulse. The latter is crucial for timing and has not been addressed in previous THGEM studies. Since the induced signal in a conventional THGEM operated with an induction gap above 1~mm is largely determined by the motion of electrons in the induction region, a smaller induction gap is expected to result in a faster current signal and improved timing performance.

In previous works, THGEM detectors were operated with small induction gaps down to 400~$\upmu$m in Ne:CF$_4$ (95:5)~\cite{Coimbra_2013} and down to 500~$\upmu$m in Ne:CH$_4$ (95:5) and Ar:CH$_4$ (80:20)~\cite{coimbra2017characterization} at different field configurations, resulting in high gain due to charge amplification in the induction gap. In the present work, the detector is operated in a mildly quenched Ar:CO$_2$ (93:7) mixture with an induction gap as small as 100~$\upmu$m. This would be challenging using a metallic anode because of the intensity of occurring electric discharges even at voltages lower than the operating voltages used here (see section~\ref{discharge_quenching}).

This paper is organized as follows: Section~\ref{sec:exp_setup_methodology} describes the experimental setup and methodology, Section~\ref{sec:results} presents the results, and Section~\ref{sec:discussion} provides the discussion.

\section{Experimental setup and methodology}
\label{sec:exp_setup_methodology}

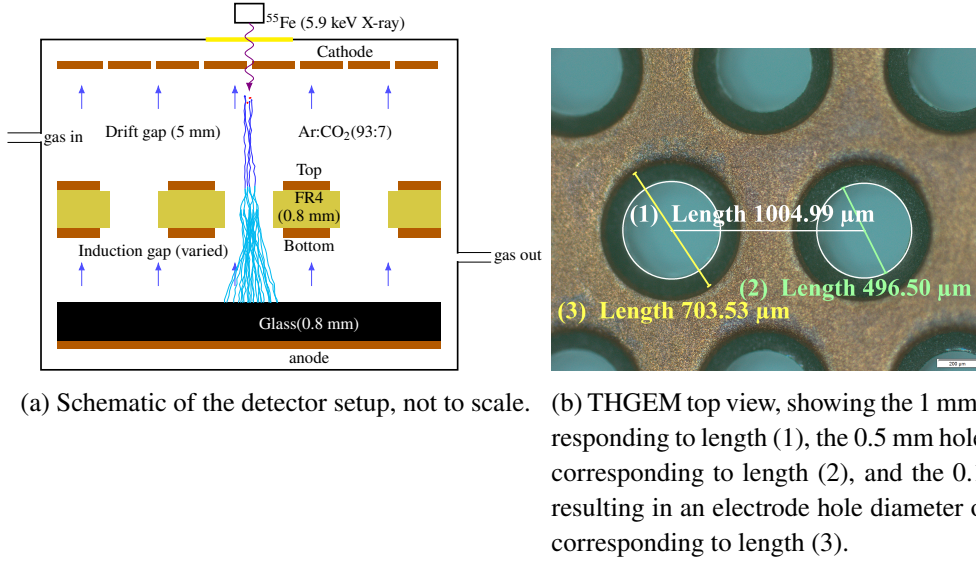
\begin{figure}[htbp]
\centering
\begin{subfigure}[t]{0.47\textwidth}
\begin{tikzpicture}[
  >=Latex,
  layer/.style={draw},
  cu/.style={fill=orange!70!black, draw=none},
  fr4/.style={fill=yellow!80!black, draw=none},
  field/.style={blue!70, -Latex},
  electrode/.style={fill=orange!70!black, draw=none},
  glass/.style={fill=black, draw=none},
  wire/.style={black, -Latex, line width=1.0pt},
  box/.style={draw=black, line width=1.0pt},
  sbox/.style={draw=black, line width=1.0pt},
  ray/.style={violet, line width=0.5pt, decorate, decoration={snake}},
  kapton/.style={fill=yellow, draw=yellow, line width=0pt},
  pipe/.style={draw=black, line width=0.6pt},
  clean/.style={draw=white, line width=4pt},
  dimtxt/.style={font=\fontsize{4}{4.5}\selectfont, inner sep=1.5pt,color=black},
  drift/.style={blue!80, line width=0.3pt, decorate,
                decoration={random steps, segment length=3pt, amplitude=0.6pt}},
  ion/.style={red, fill=red, circle, inner sep=0.25pt},
  electron/.style={blue!80, fill=blue, circle, inner sep=0.25pt},
  avalanche/.style={cyan!80, line width=0.1pt, decorate,
                    decoration={random steps, segment length=3pt, amplitude=0.6pt}}
]

\def\W{8.0}
\def\Hcu{0.20}
\def\Hfr{0.80}
\def\gapD{2.5}
\def\gapI{1.0}
\def\topY{0}
\def\Hcath{0.15}
\def\Hanod{0.15}
\def\Hglass{0.8}

\coordinate (ThTop)  at (0,\topY);
\coordinate (CuTopB) at (0,\topY-\Hcu);
\coordinate (FrB)    at (0,\topY-\Hcu-\Hfr);
\coordinate (CuBotB) at (0,\topY-\Hcu-\Hfr-\Hcu);

\pgfmathsetmacro{\yCathTop}{\topY+\gapD}
\pgfmathsetmacro{\yCathBot}{\yCathTop-\Hcath}
\fill[electrode] (0,\yCathTop) rectangle (\W,\yCathBot);

\def\dxCath{0.05}
\def\eps{0.03}
\foreach \x in {1.0,2.0,3.0,4.0,5.0,6.0,7.0,8.0} {
  \fill[white] (\x-\dxCath,\yCathTop+\eps) rectangle (\x+\dxCath,\yCathBot-\eps);
}

\draw[cu]  (0,\topY) rectangle (\W,\topY-\Hcu);
\draw[fr4] (0,\topY-\Hcu) rectangle (\W,\topY-\Hcu-\Hfr);
\draw[cu]  (0,\topY-\Hcu-\Hfr) rectangle (\W,\topY-\Hcu-\Hfr-\Hcu);

\def\dxCu{0.7}
\def\dxFr{0.5}
\foreach \x in {1.6,4.0,6.4} {
  \fill[white] (\x-\dxCu,\topY+0.1) rectangle (\x+\dxCu,\topY-\Hcu);
  \fill[white] (\x-\dxFr,\topY-\Hcu+0.1) rectangle (\x+\dxFr,\topY-\Hcu-\Hfr-0.1);
  \fill[white] (\x-\dxCu,\topY-\Hcu-\Hfr) rectangle (\x+\dxCu,\topY-\Hcu-\Hfr-\Hcu-0.1);
}

\pgfmathsetmacro{\yAnodTop}{\topY-\Hcu-\Hfr-\Hcu-\gapI-1.15}
\pgfmathsetmacro{\yAnodBot}{\yAnodTop-\Hanod}
\fill[electrode] (0,\yAnodTop) rectangle (\W,\yAnodBot);

\pgfmathsetmacro{\yGlassTop}{\yAnodTop+\Hglass}
\fill[glass] (0,\yGlassTop) rectangle (\W,\yAnodTop);

\foreach \x in {0.8,2.4,4.0,5.6,7.2} {
  \draw[field] (\x-0.3,\topY+1.5) -- (\x-0.3,\topY+2.0);
  \draw[field] (\x-0.3,\topY-\Hcu-\Hfr-\Hcu-\gapI) --
               (\x-0.3,\topY-\Hcu-\Hfr-\Hcu-\gapI+0.5);
}

\node[dimtxt] at (\W/2+2.0,\topY+1.0) {\small Ar:CO\textsubscript{2}(93:7)};
\node[dimtxt] at (\W/2-1.8,\topY+1.0) {\small Drift gap (5~mm)};
\node[dimtxt] at (\W/2-2.0,\topY-1.50) {\small Induction gap (varied)};
\node[dimtxt] at (\W/2+2.0,\topY+2.7) {\small Cathode};
\node[dimtxt] at (\W/2+1.250,\topY+0.2) {\small Top};
\node[dimtxt] at (\W/2+1.250,\topY-0.35) {\small FR4};
\node[dimtxt] at (\W/2+1.250,\topY-0.75) {\small (0.8~mm)};
\node[dimtxt] at (\W/2+1.250,\topY-1.35) {\small Bottom};
\node[dimtxt] at (\W/2+1.250,\topY-3.0) {\small \color{white}{Glass(0.8~mm)}};
\node[dimtxt] at (\W/2+1.250,\topY-3.70) {\small anode};

\def\padX{0.35}
\def\padY{0.45}
\coordinate (GasTL) at (-\padX, \yCathTop+\padY);
\coordinate (GasBR) at (\W+\padX, \yAnodBot-\padY);
\draw[box] (GasTL) rectangle (GasBR);

\pgfmathsetmacro{\xWinL}{\W/2-0.9}
\pgfmathsetmacro{\xWinR}{\W/2+0.9}
\pgfmathsetmacro{\yWinTop}{\yCathTop+\padY}
\fill[kapton] (\xWinL,\yWinTop-0.035) rectangle (\xWinR,\yWinTop+0.035);

\coordinate (SrcC) at (\W/2, \yCathTop+\padY+0.55);
\draw[sbox] ($(SrcC)+(-0.30,0.20)$) rectangle ($(SrcC)+(0.30,-0.20)$);
\node[right,dimtxt] at ($(SrcC)+(0.30,-0.2)$) {\small $^{55}$Fe (5.9 keV X-ray)};

\draw[ray, -Latex] ($(SrcC)+(0,-0.20)$) -- (\W/2, \yCathTop-0.64);

\pgfmathsetmacro{\yPipe}{0.5*(\yCathTop+\yAnodBot)}
\draw[pipe] (-\padX,\yPipe+1.5) -- (-\padX-0.70,\yPipe+1.5);
\draw[pipe] (-\padX,\yPipe+1.3) -- (-\padX-0.70,\yPipe+1.3);
\draw[clean] (-\padX,\yPipe+1.322) -- (-\padX,\yPipe+1.478);
\node[right,dimtxt] at (-\padX,\yPipe+1.4) {\small gas in};

\draw[pipe] (\W+\padX,\yPipe-1.0) -- (\W+\padX+0.70,\yPipe-1.0);
\draw[pipe] (\W+\padX,\yPipe-1.2) -- (\W+\padX+0.70,\yPipe-1.2);
\draw[clean] (\W+\padX,\yPipe-1.178) -- (\W+\padX,\yPipe-1.022);
\node[right,dimtxt] at (\W+\padX+0.70,\yPipe-1.1) {\small gas out};

\def\nElectrons{4}
\def\holeX{4.0}
\coordinate (IntPoint) at (\W/2, \yCathTop-0.8);

\foreach \i in {1,...,\nElectrons}{
  \node[ion] at ($(IntPoint)+(rand*0.1, rand*0.1)$) {};
  \node[electron] (e\i) at ($(IntPoint)+(rand*0.1, rand*0.1)$) {};
}

\foreach \i in {1,...,\nElectrons}{
  \pgfmathsetmacro{\targetX}{\holeX + rand*0.15}
  \pgfmathsetmacro{\targetY}{\topY -0.1- rnd*0.1}
  \draw[drift] (e\i.center) .. controls ++(0,-0.8) and ++(0,0.8) ..
      (\targetX,\targetY) node(startAvalanche\i) {};

  \foreach \b in {1,2}{
    \coordinate (midPoint\i\b) at (\targetX + rand*0.2, \topY - 0.6);
    \draw[avalanche] (startAvalanche\i.center)
      .. controls ++(0,-0.1) and ++(0,0.1) .. (midPoint\i\b);

    \foreach \shower in {1,2,3}{
      \draw[avalanche] (midPoint\i\b)
        .. controls ++(0,-0.2) and ++(0,0.2) ..
        (\holeX + rand*0.6, \yGlassTop);
    }
  }
}

\pgfmathsetmacro{\yCathMid}{0.5*(\yCathTop+\yCathBot)}
\pgfmathsetmacro{\yTopMid}{\topY-\Hcu/2}
\pgfmathsetmacro{\yBotMid}{\topY-\Hcu-\Hfr-\Hcu/2}
\pgfmathsetmacro{\yAnodMid}{0.5*(\yAnodTop+\yAnodBot)}
\def\xBus{\W+1.25}



\end{tikzpicture}
\caption{Schematic of the detector setup, not to scale.}
\label{detector_setup}
\end{subfigure}
\begin{subfigure}[t]{0.47\textwidth}
\includegraphics[width=0.8\linewidth]{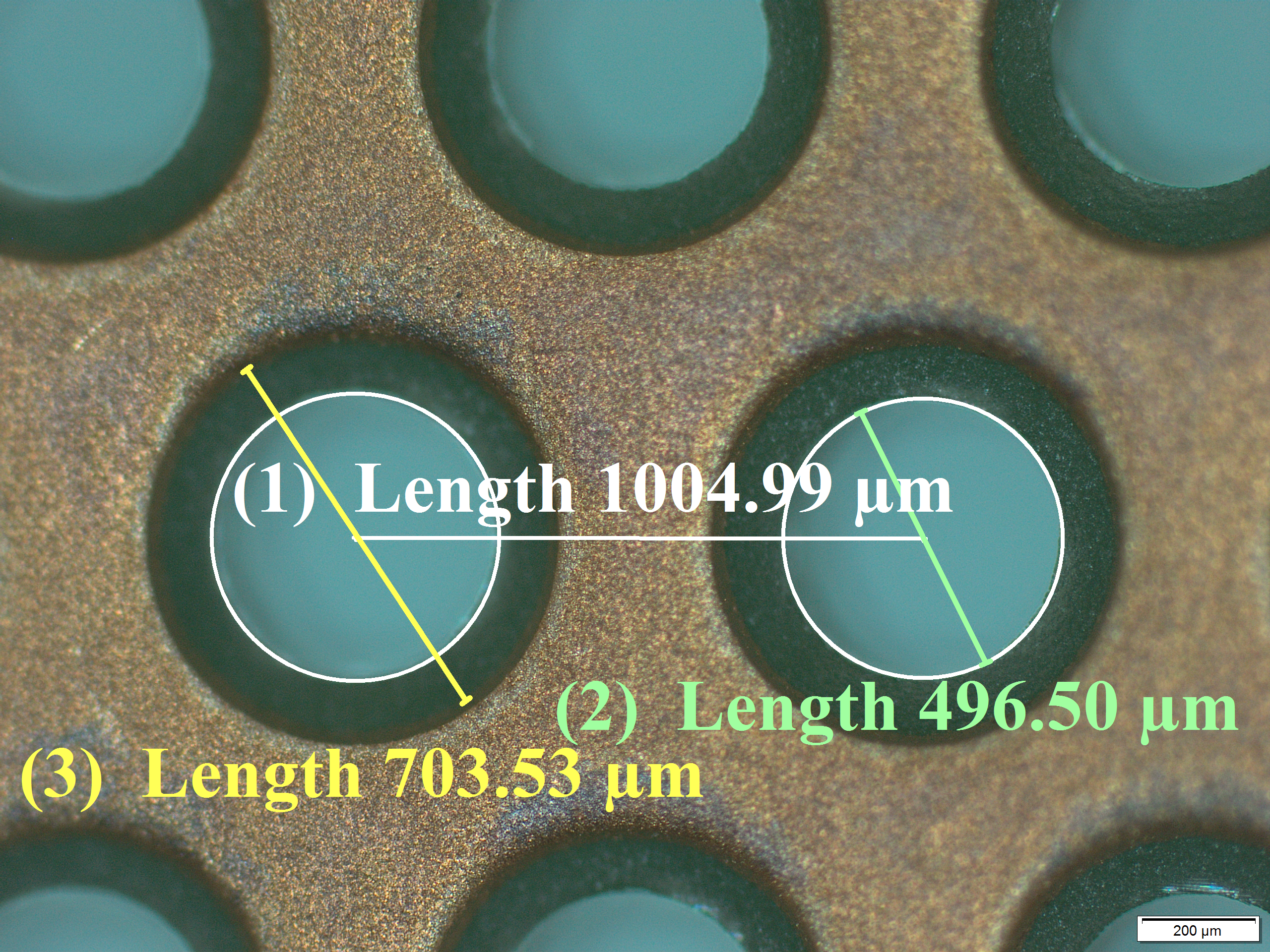}
\caption{THGEM top view, showing the 1~mm pitch, corresponding to length (1), the 0.5~mm hole diameter, corresponding to length (2), and the 0.1~mm rim, resulting in an electrode hole diameter of 0.7~mm, corresponding to length (3).}
\label{thgem_topview}
\end{subfigure}
\caption{Detector setup schematic(not to scale) and microscope image for the THGEM.}
\end{figure}

The schematic of the detector assembly is shown in Figure~\ref{detector_setup}. The detector consists of a THGEM, a cathode electrode, and a resistive anode, each with an active area of 2.5~cm $\times$ 2.5~cm. The THGEM, shown in Figure~\ref{thgem_topview}, is fabricated from a 0.8~mm-thick FR4 plate clad with $\sim$ 90 $\mu$m copper on both sides. It is perforated with 0.5~mm-diameter holes arranged in a hexagonal pattern with a pitch of 1~mm. A 0.1~mm rim is formed around each hole by copper etching. The cathode, made of a stainless-steel mesh to allow the transmission of soft X-rays, is positioned 5~mm above the THGEM, defining the drift gap. The anode consists of a 0.8~mm-thick resistive glass plate electrically coupled to a copper readout pad through conductive tape\footnote{3M 9707\texttrademark}. The glass has a bulk resistivity of approximately $10^{10}~\Omega\cdot$cm~\cite{WANG2010151}.

The distance between the glass surface and the bottom face of the THGEM insulator defines the induction gap. The induction-gap thickness was varied using precision spacers to obtain gaps of 0.1, 0.2, 0.5, 1, and 2~mm. The detector assembly was enclosed in a chamber continuously flushed with an Ar:CO$_2$ (93:7) gas mixture at a flow rate of approximately 20~sccm.

\begin{figure}[htbp]
    \centering
    \begin{subfigure}[b]{0.48\textwidth}
    \includegraphics[width=\linewidth]{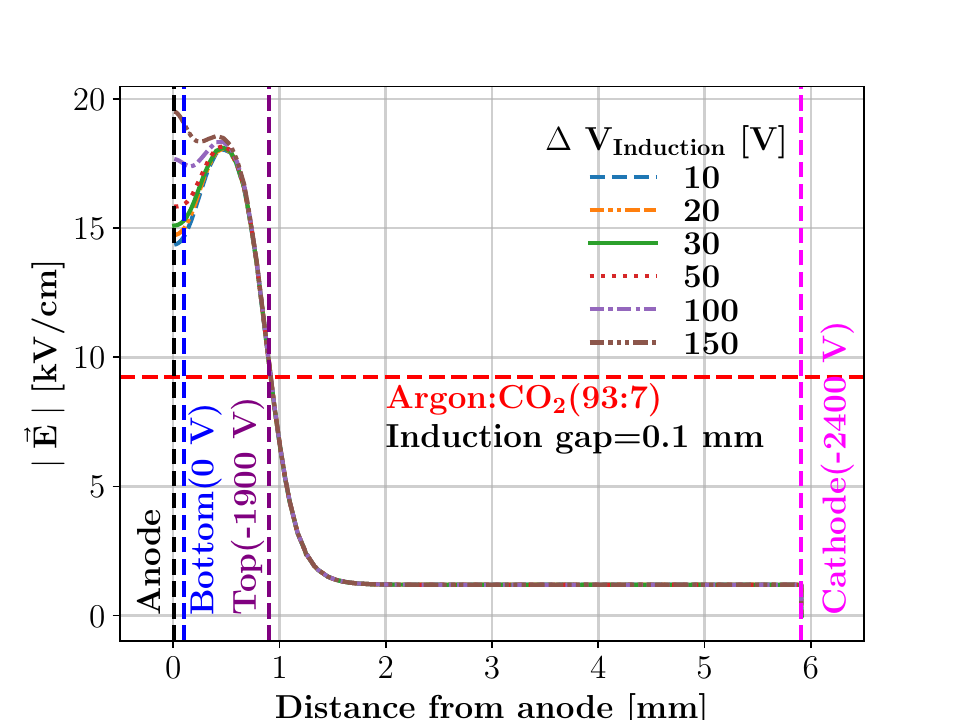}
    \caption{Drift field = 1~kV/cm, \DV{THGEM} = 1900~V, for different \DV{induction} values, as shown in the legend, for a 0.1~mm induction gap.}
    \label{electric_field_100_um}
    \end{subfigure}
    \hfill
    \begin{subfigure}[b]{0.48\textwidth}
    \includegraphics[width=\linewidth]{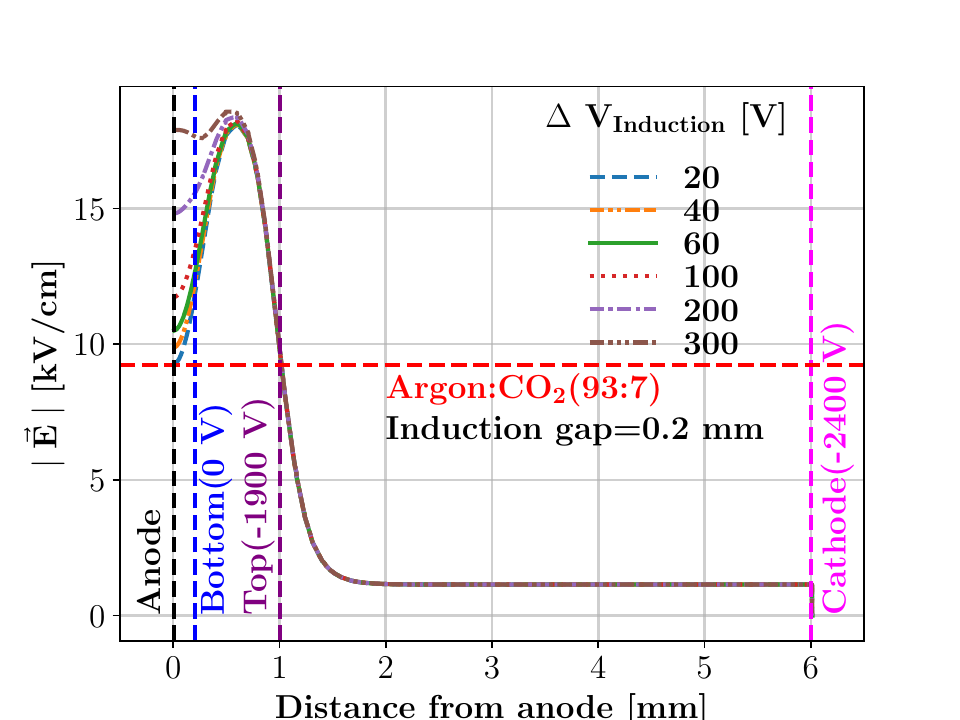}
    \caption{Drift field = 1~kV/cm, \DV{THGEM} = 1900~V, for different \DV{induction} values, as shown in the legend, for a 0.2~mm induction gap.}
    \label{electric_field_200_um}
    \end{subfigure}\\
    \begin{subfigure}[b]{0.48\textwidth}
    \includegraphics[width=\linewidth]{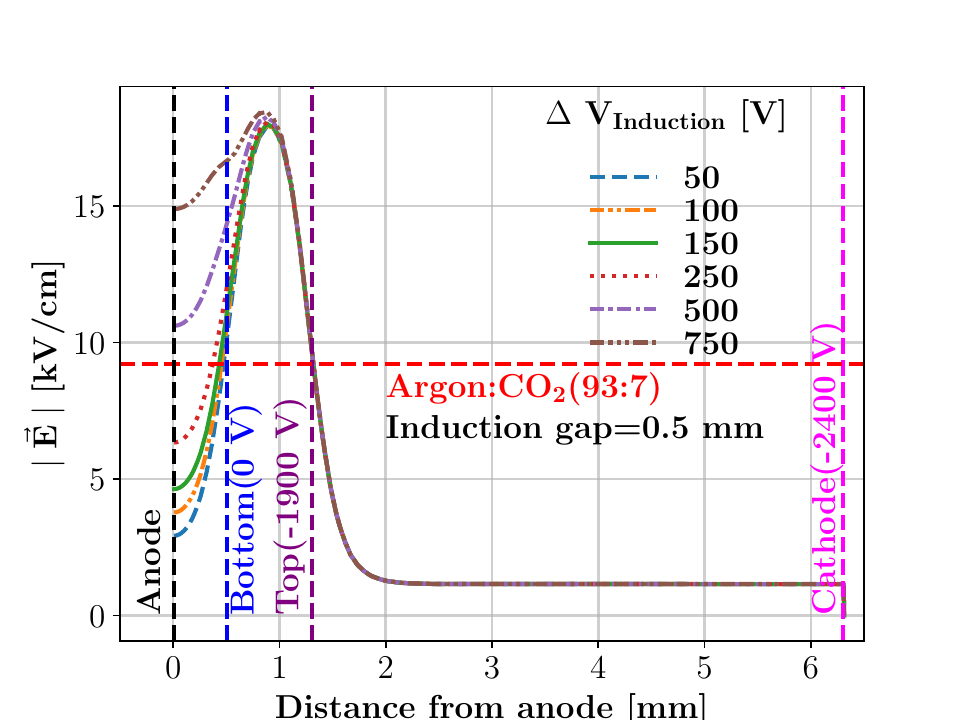}
    \caption{Drift field = 1~kV/cm, \DV{THGEM} = 1900~V, for different \DV{induction} values, as shown in the legend, for a 0.5~mm induction gap.}
    \label{electric_field_500_um}
    \end{subfigure}
    \hfill
    \begin{subfigure}[b]{0.48\textwidth}
    \includegraphics[width=\linewidth]{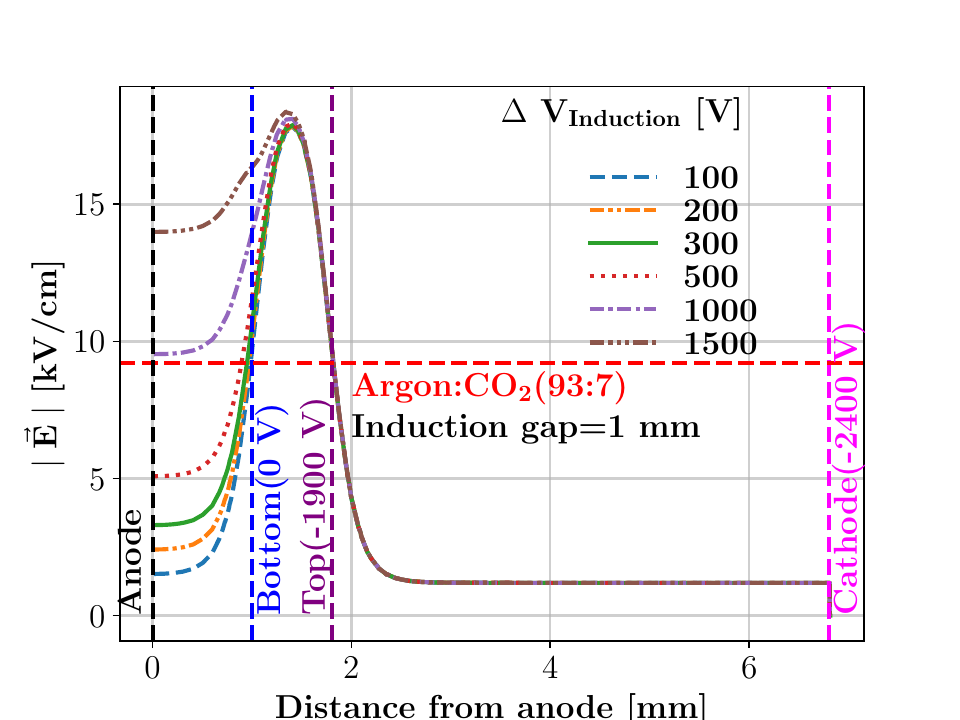}
    \caption{Drift field = 1~kV/cm, \DV{THGEM} = 1900~V, for different \DV{induction} values, as shown in the legend, for a 1.0~mm induction gap.}
    \label{electric_field_1_mm}
    \end{subfigure}\\
    \begin{subfigure}[b]{0.48\textwidth}
    \includegraphics[width=\linewidth]{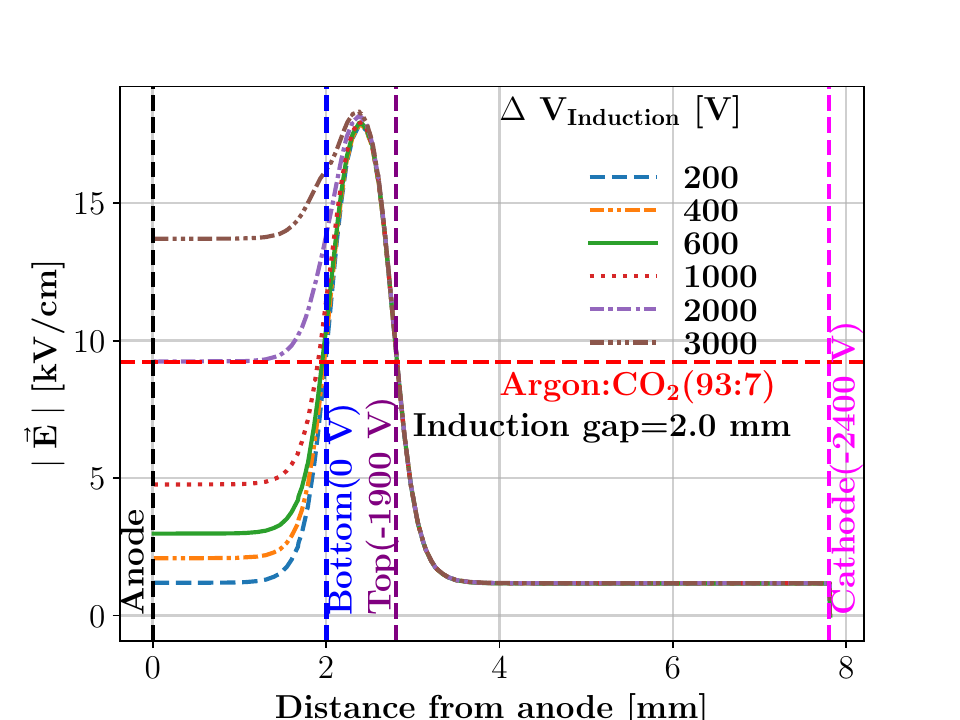}
    \caption{Drift field = 1~kV/cm, \DV{THGEM} = 1900~V, for different \DV{induction} values, as shown in the legend, for a 2.0~mm induction gap.}
    \label{electric_field_2_mm}
    \end{subfigure}
    \caption{THGEM electric field for different induction gaps, simulated with \texttt{COMSOL} \cite{comsol_multiphysics_6_2}. The electric-field norm, in kV/cm, is evaluated along a line passing through the center of the hole. The red horizontal dashed line corresponds to the Townsend avalanche threshold in Ar:CO$_2$ (93:7).}
    \label{electric_field}
\end{figure}

Primary electrons are produced in the drift gap by 5.9~keV X-rays from a $^{55}$Fe source or by cosmic muons, and are drifted toward the multiplication region by a drift field of 1~kV/cm. They are then multiplied through Townsend avalanche processes within the THGEM holes and, depending on the specific field configuration, possibly also in the induction gap. Figure~\ref{electric_field} shows the THGEM electric field for typical voltage configurations, simulated with \texttt{COMSOL}. It should be noted that the actual field, in particular for small induction gap values, is higher than that of a parallel plate configuration. For example, a voltage difference of 40~V across a 0.2~mm induction gap would correspond to 2~kV/cm in the parallel-plate approximation, but it results in an actual field of about 10~kV/cm, exceeding the threshold for Townsend avalanche development\footnote{Townsend coefficient minus attachment coefficient greater than zero (simulated with \texttt{Magboltz})} in Ar:CO$_2$ (93:7).

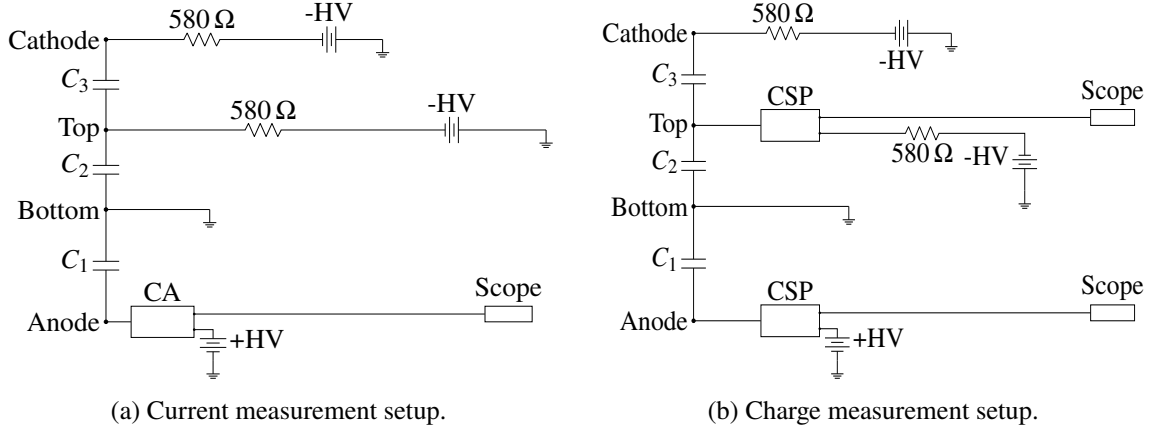
\begin{figure}[htbp]
\centering
\begin{subfigure}[t]{0.48\linewidth}
\begin{circuitikz}[american, transform shape]


\coordinate (B0) at (-5.8,-1.15);
\coordinate (B1) at (-5.8,2.45);
\coordinate (B2) at (-5.8,5.0);
\coordinate (B3) at (-5.8,7.9);

\draw (B0) node[circ, label=left:\Huge Anode]{};
\draw (B1) node[circ, label=left:\Huge Bottom]{};
\draw (B2) node[circ, label=left:\Huge Top]{};
\draw (B3) node[circ, label=left:\Huge Cathode]{};

\draw (B0) to[C,l=\Huge $C_1$] (B1);
\draw (B1) to[C,l=\Huge $C_2$] (B2);
\draw (B2) to[C,l=\Huge $C_3$] (B3);

\node[draw, rectangle, minimum width=2.0cm, minimum height=1.0cm,
      label=above:{\Huge CA}] (CA1) at (-4.0,-1.15) {};
\draw (B0) -- (CA1.west);

\coordinate (CA1outU) at ($(CA1.east)+(0,0.25)$);
\coordinate (CA1outL) at ($(CA1.east)+(0,-0.25)$);

\fill (CA1outU) circle (1.2pt);
\fill (CA1outL) circle (1.2pt);

\node[draw, rectangle, minimum width=1.5cm, minimum height=0.5cm,
      label=above:{\Huge Scope}] (ScA) at ($(CA1outU)+(10.0,0)$) {};

\draw (CA1outU) -- (ScA.west);
\draw (CA1outL) -- ++(0.6,0)
      to[battery,l=\Huge +HV] ++(0,-1.0)
      node[ground]{};

\draw (B1) -- (-2.5,2.45) node[ground]{};

\draw (B2)
  to[R,l=\Huge $580\,\Omega$] ++(10.0,0)
  to[battery,invert,l=\Huge -HV] ++(2.0,0)
  -- ++(2.0,0)
  node[ground]{};

\draw (B3)
  to[R,l=\Huge $580\,\Omega$] ++(6.1,0)
  to[battery,invert,l=\Huge -HV] ++(2.0,0)
  -- ++(0.7,0)
  node[ground]{};

\end{circuitikz}
\caption{Current measurement setup.}
\label{current_measurement_setup}    
\end{subfigure}
\hfill
\begin{subfigure}[t]{0.48\linewidth}
\begin{circuitikz}[american, transform shape]

\coordinate (A0) at (-12.3,-1.15);
\coordinate (A1) at (-12.3, 2.45);
\coordinate (A2) at (-12.3, 5.00);
\coordinate (A3) at (-12.3, 7.90);

\draw (A0) node[circ, label=left:\Huge Anode]{};
\draw (A1) node[circ, label=left:\Huge Bottom]{};
\draw (A2) node[circ, label=left:\Huge Top]{};
\draw (A3) node[circ, label=left:\Huge Cathode]{};

\draw (A0) to[C,l=\Huge $C_1$] (A1);
\draw (A1) to[C,l=\Huge $C_2$] (A2);
\draw (A2) to[C,l=\Huge $C_3$] (A3);

\node[draw, rectangle,
      minimum width=2.0cm,
      minimum height=1.0cm,
      label=above:{\Huge CSP}] (CSP1) at (-9.0,-1.15) {};

\draw (A0) -- (CSP1.west);

\coordinate (CSP1outU) at ($(CSP1.east)+(0,0.25)$);
\coordinate (CSP1outL) at ($(CSP1.east)+(0,-0.25)$);

\fill (CSP1outU) circle (1.2pt);
\fill (CSP1outL) circle (1.2pt);

\node[draw, rectangle,
      minimum width=1.5cm,
      minimum height=0.5cm,
      label=above:{\Huge Scope}] (Scope1) at ($(CSP1outU)+(10.0,0)$) {};

\draw (CSP1outU) -- (Scope1.west);

\draw (CSP1outL) -- ++(0.6,0)
      to[battery,l=\Huge +HV] ++(0,-1.0)
      node[ground]{};

\draw (A1) -- (-7.0,2.45) node[ground]{};

\node[draw, rectangle,
      minimum width=2.0cm,
      minimum height=1.0cm,
      label=above:{\Huge CSP}] (CSP2) at (-9.0,5.0) {};

\draw (A2) -- (CSP2.west);

\coordinate (CSP2outU) at ($(CSP2.east)+(0,0.25)$);
\coordinate (CSP2outL) at ($(CSP2.east)+(0,-0.25)$);

\fill (CSP2outU) circle (1.2pt);
\fill (CSP2outL) circle (1.2pt);

\node[draw, rectangle,
      minimum width=1.5cm,
      minimum height=0.5cm,
      label=above:{\Huge Scope}] (Scope2) at ($(CSP2outU)+(10.0,0)$) {};

\draw (CSP2outU) -- (Scope2.west);

\draw (CSP2outL)
      to[R,l_=\Huge $580\,\Omega$] ++(7.0,0)
      to[battery,invert,l_=\Huge -HV] ++(0,-1.8)
      node[ground]{};

\draw (A3)
      to[R,l=\Huge $580\,\Omega$] ++(6.1,0)
      to[battery,invert,l_=\Huge -HV] ++(2.0,0)
      -- ++(0.7,0)
      node[ground]{};

\end{circuitikz}
\caption{Charge measurement setup.}
\label{charge_measurement_setup}    
\end{subfigure}
\caption{Detector connections to the high-voltage bias and readout electronics.}
\label{high_voltage_bias}
\end{figure}

Figure~\ref{high_voltage_bias} shows a schematic of the detector connections to the high-voltage bias and readout electronics, where $C_1$, $C_2$, and $C_3$ are the effective capacitances between the anode and bottom electrode, the bottom and top electrodes, and the top electrode and cathode, respectively. The cathode was always biased with a negative high voltage through a 580~$\Omega$ resistor. Together with the capacitance of the HV cable, this resistance forms a low-pass RC filter that mitigates high frequency noise from the high-voltage lines.

The bottom electrode of the THGEM was grounded using a short copper-nickel braid connected to the chamber body, providing a low-impedance path for the current induced on the bottom electrode. If the bottom electrode is not firmly grounded, current induced on it may capacitively couple to the anode through the induction gap and contribute to the measured signal. The strength of this coupling is governed by the capacitive impedance of the induction gap,
\begin{equation}
    Z_C = \frac{1}{j\omega C_1},
\end{equation}
where $\omega$ is the angular frequency of the signal. Since $C_1$ increases as the induction gap decreases, the corresponding impedance becomes smaller for narrow gaps, allowing high frequency components of the induced current to be transferred more efficiently across the gap. It was also found that this grounding scheme suppresses high frequency oscillations and reflections in the signal line.

The THGEM and anode electrodes were connected in one of two distinct configurations, either for the measurement of fast induced current waveforms, as shown in Figure~\ref{current_measurement_setup}, or for the measurement of integrated induced current (i.e. charge) waveforms, as shown in Figure~\ref{charge_measurement_setup}.

\paragraph{Current waveforms}
The THGEM-top electrode is biased in a similar way to the cathode. The anode is biased with a positive high voltage through a \texttt{CIVIDEC C2-TCT} current amplifier (CA), providing an amplification factor of $\sim$100 and a fast step-pulse rise time of $\sim$330 ps. 

\paragraph{Charge waveforms}
A multi-channel \texttt{NAICAM DCB209} charge sensitive preamplifier (CSP) was used to read out both the THGEM-top electrode and the anode. The THGEM-top electrode is biased with a negative high voltage through an additional 580~$\Omega$ resistor before the CSP, while the anode is biased with a positive high voltage, as shown in Figure~\ref{charge_measurement_setup}. The anode was not biased through a resistor in order to maintain a low-impedance path, ensuring that the induced current remains on the anode rather than flowing to the grounded bottom electrode.

All waveforms were digitized using a \texttt{Tektronix MSO46} oscilloscope at a sampling rate of 3.125~GS/s with a bandwidth of 250~MHz. The current signals were coupled through 50~$\Omega$ DC, while the charge signals were coupled through 1~M$\Omega$ AC.

The properties of the current and charge signals were studied as a function of different induction-gap values. The induction gap configuration giving the best performance was further investigated under different amplification and induction-field conditions.

\subsection{Waveform analysis}

Baseline correction was performed on each digitized waveform by subtracting the mean amplitude measured in the pre-signal region. Waveform features of interest were then extracted from each waveform and stored in histograms. To determine the characteristic values of the measured quantities, each distribution was fitted using a two-step Gaussian procedure: the initial fit was performed within the distribution mean $\pm1$ standard deviation, and the second fit was performed within the fitted mean $\pm1$ standard deviation obtained from the initial fit. In the following plots, the markers represent the resulting fit means. The error bars represent the error on the fitted mean. The shaded bands indicate $\pm$1 standard deviation of the fitted Gaussian.

All reported quantities were derived from data sets containing approximately 4000 recorded waveforms.

\paragraph{Current waveforms}
The digitized current waveforms were first processed with a Fourier filter~\cite{2020SciPy-NMeth} using a cutoff frequency of 0.1~GHz to suppress high frequency noise. The current-waveform properties studied were: the peak amplitude, rise time, pulse duration, and fast charge component. The \textit{peak amplitude} is the maximum amplitude of the signal with respect to the baseline. The \textit{rise time} is defined as the time required for the signal to increase from 20\% to 90\% of the peak amplitude along the rising edge. The 20\% threshold was chosen to avoid small amplitude fluctuations. The \textit{pulse duration} is defined as the time interval between the points at which the signal reaches 20\% of its peak amplitude on the rising and falling edges. The \textit{fast-component charge} is obtained by integrating the signal over the \textit{pulse duration} defined above. The integral is then converted to charge units using the amplifier gain factor and the 50~$\Omega$ impedance of the oscilloscope input, as follows:
\begin{equation}
Q = \frac{1}{A \cdot (50\,\Omega)} \sum_{i=1}^{N} V(t_i)\,\cdot\Delta t
\end{equation}
where $V(t_i)$ is the voltage measured by the oscilloscope at time $t_i$, $\Delta t$ is the sampling interval, $A\approx100$ is the gain of the current amplifier, and $50\,\Omega$ is the oscilloscope impedance. 

\paragraph{Charge waveforms}
The charge-waveform properties studied were: the anode signal charge, the anode signal rise time, the THGEM-top charge, and the anode/THGEM-top charge ratio. The \textit{anode charge} and \textit{THGEM-top charge} are the maximum values of the anode and THGEM-top waveforms, respectively. They represent the charge collected at each electrode. The \textit{anode/THGEM-top charge ratio} is the ratio of the charge collected at the anode with respect to the charge collected at the top electrode. Under the assumption that all avalanche ions are collected at the THGEM-top electrode, this ratio approximately represents the fraction of the total avalanche electrons reaching the anode. The \textit{anode signal rise time} is defined as the time required for the anode signal to increase from 10\% to 90\% of the total charge.

\subsection{Time-resolution study}

After identifying the induction gap and field configuration that provide current signals with the largest peak amplitude and fastest rise time, the THGEM time resolution with current readout was studied detecting cosmic muons. Two 50 $\times$ 50 $\times$ 10 mm$^3$ scintillators\footnote{\texttt{Shalom EO SP101}} coupled to photomultiplier tubes (PMTs)\footnote{\texttt{Beijing Hamamatsu CR185}} were used to provide a precise time reference. The resolution of each scintillator was measured to be approximately 300~ps. The scintillator signals and the THGEM anode current signals were digitized at a sampling rate of 12.5~GS/s and bandwidth of 250~MHz. All signals were then interpolated and their time was defined at 20\% of the signal peak amplitude. Defining
$t_1$ and $t_2$- time measured with two scintillators, the reference time is
\begin{equation}
 t_{\text{ref}} = \frac{t_1 + t_2}{2},
\end{equation}

For the THGEM signal, only the first peak, corresponding to the closest primary ionization cluster was considered.
Finally, the time resolution of the THGEM is obtained using equation~\ref{thgem_t_res}:
\begin{equation}
 \sigma_{\text{THGEM}} = \sqrt{\text{Var}(t_{\text{THGEM}} - t_{\text{ref}})} ,
 \label{thgem_t_res}    
\end{equation}

\section{Results}
\label{sec:results}

In the following, the parallel-plate-equivalent field values are used for the different induction-field configurations.

\subsection{Study of induction gap and field configurations}
\label{sec:Study_of_induction_gap_and_field_configurations}

Current and charge signal waveforms were acquired for different induction-gap values at the same equivalent parallel-plate induction field of 2~kV/cm. The value \DV{THGEM} = 1900~V was chosen because it was the highest \DV{THGEM} at which the detector could be operated without continuous discharges. Signals of a Resistive Plate WELL (RPWELL)~\cite{rubin2013first} comprising a single sided copper-clad electrode with similar geometry are also shown.

\paragraph{Current signals}

\begin{figure}[htbp]
    \centering
    \begin{subfigure}[t]{0.48\textwidth}
    \centering
        \includegraphics[width=\linewidth]{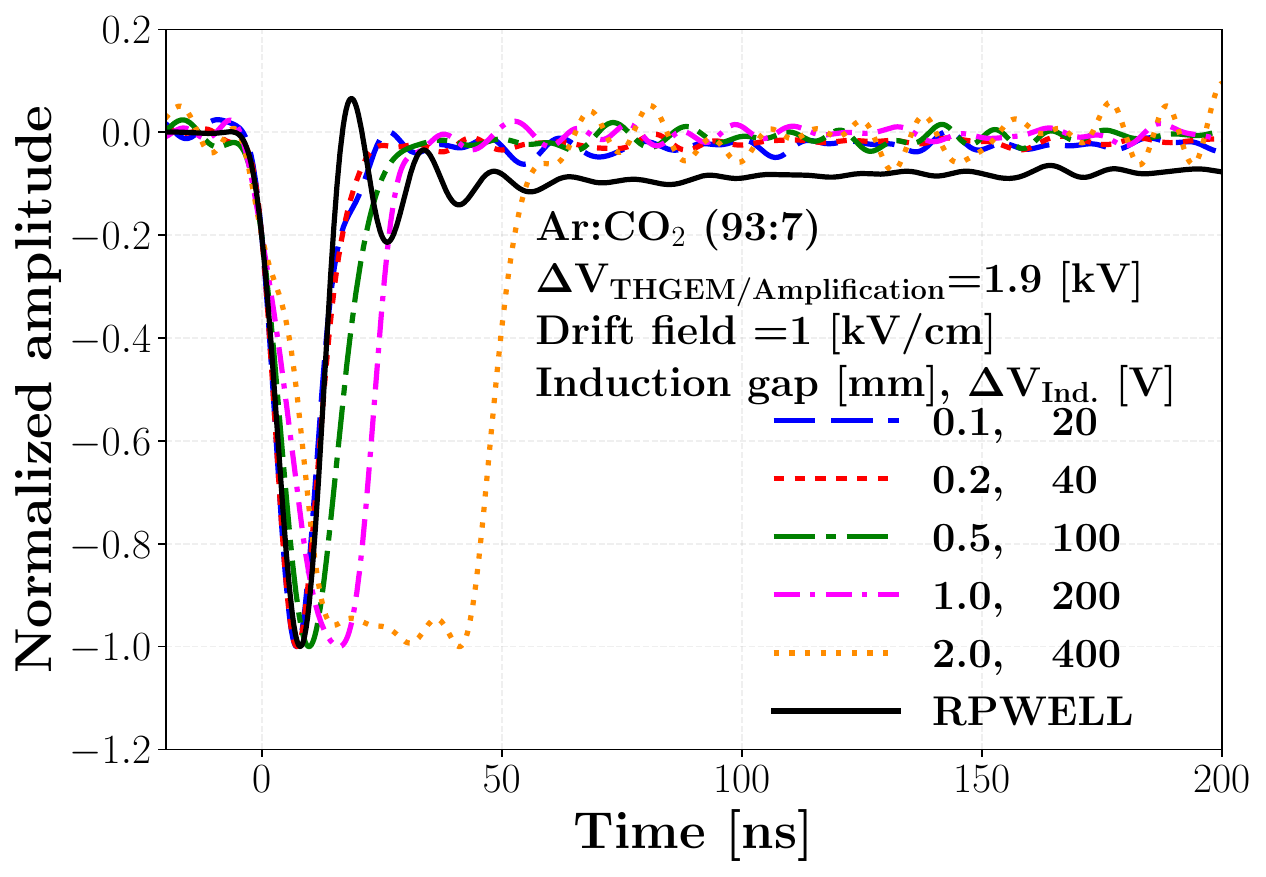}
        \caption{Typical anode current signal waveforms for RPWELL and different THGEM induction-gap values normalized to the maximum current.}
        \label{fca_all_gaps}
    \end{subfigure}
    \hfill
    \begin{subfigure}[t]{0.48\textwidth}
        \includegraphics[width=\linewidth]{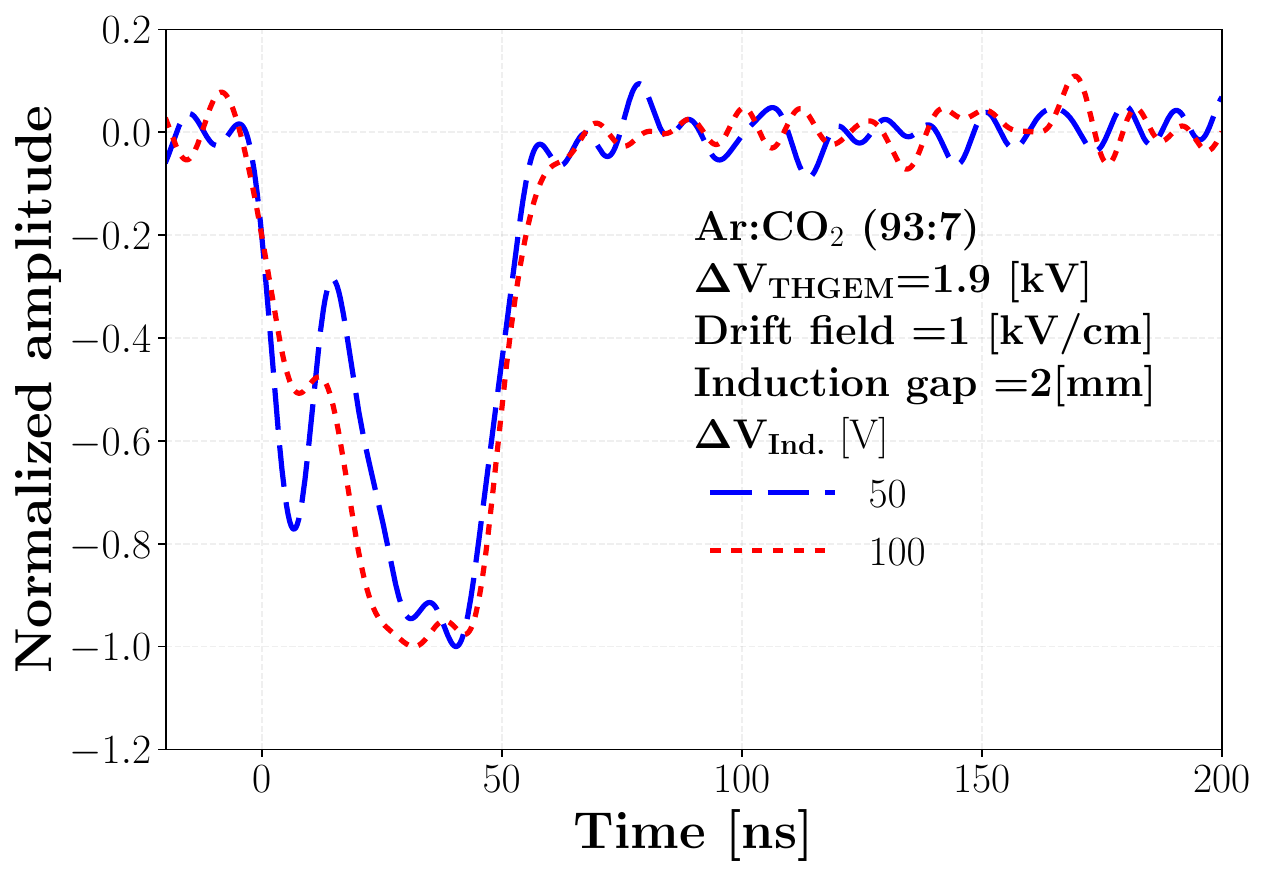}
        \caption{Typical THGEM current signal with a 2.0~mm induction gap and low induction field normalized to the maximum current.}
        \label{fca_thgem_double_feature}
    \end{subfigure}
    \caption{Typical current waveforms.}
    \label{signal_waveforms_current}
\end{figure}

Current signals with different induction gaps are shown in Figure~\ref{fca_all_gaps} for 2~kV/cm induction field parallel plate equivalent. For comparison, the waveforms were normalized to their maximum amplitude. For the THGEM configurations, the waveform becomes more peaked and symmetric as the induction gap decreases. Yet, for the RPWELL the peak is slightly broader and followed by a more pronounced ion tail.
At large induction gaps and low induction fields, a characteristic double structure is observed, as shown in Figure~\ref{fca_thgem_double_feature} for a 2~mm induction gap. It consists of an initial sharp peak, followed by a longer flat component. We attribute the first to electron motion inside or near the THGEM holes, and the second to electron motion in the induction gap. As the bias across the induction gap exceeds 300~V (corresponding to 1.5~kV/cm in parallel-plate approximation), the two components merge as seen in Figure~\ref{fca_all_gaps} and in all the following measurements.

\begin{figure}[htbp]
  \centering
  \begin{subfigure}[t]{0.48\textwidth}
    \centering
    \includegraphics[width=\textwidth]{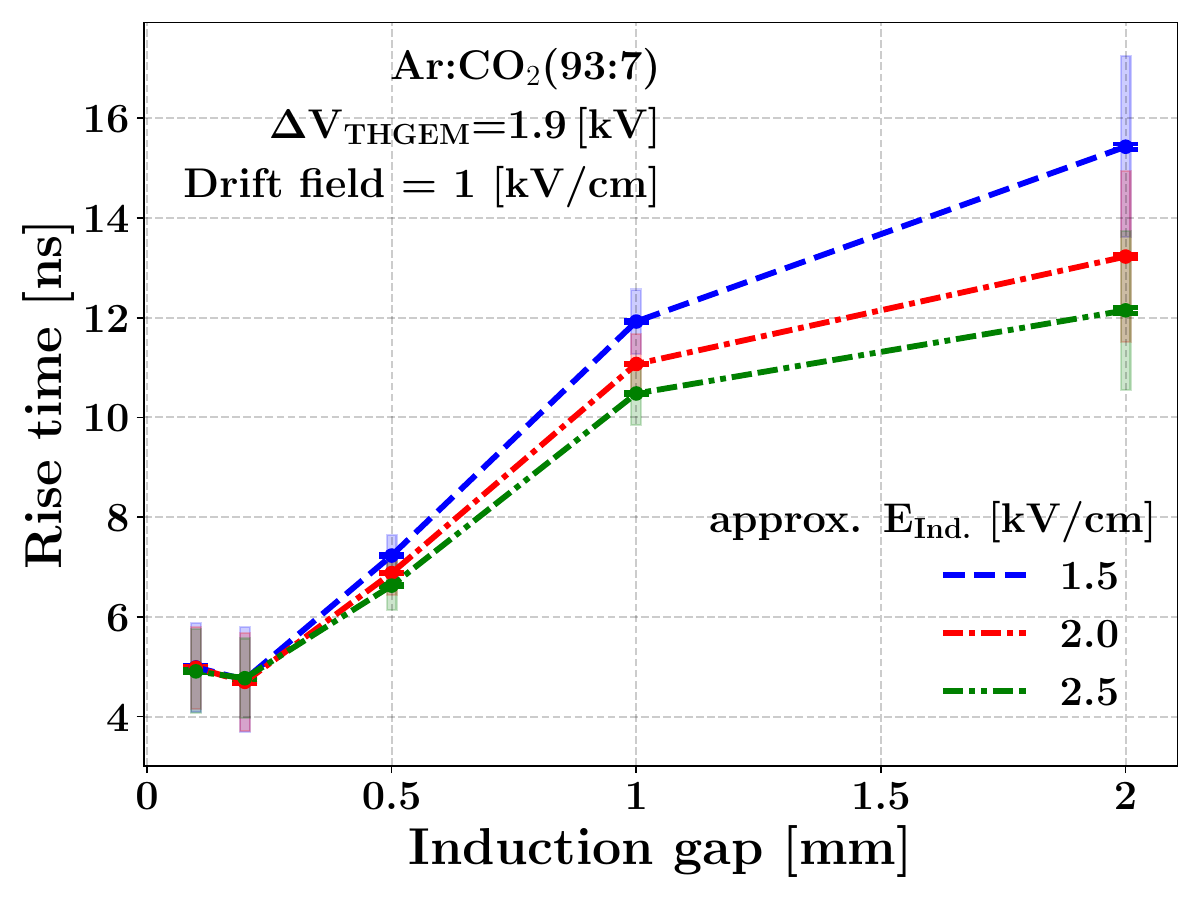}
    \caption{Rise time.}
    \label{fig_rise_time_induction_gap}
  \end{subfigure}
  \hfill
  \begin{subfigure}[t]{0.48\textwidth}
    \centering
    \includegraphics[width=\textwidth]{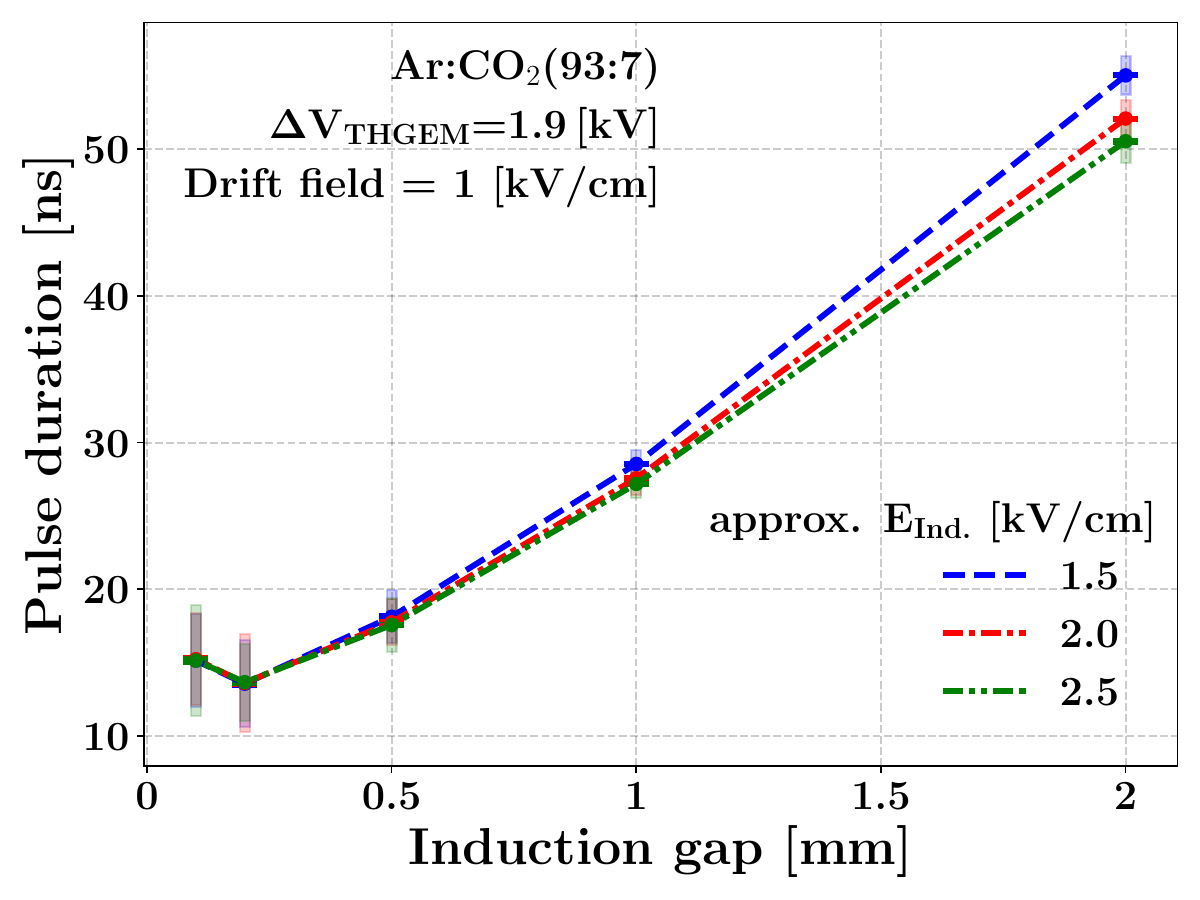}
    \caption{Pulse duration.}
    \label{fig_time_width_induction_gap}
  \end{subfigure}

  \vspace{0.5cm}

  \begin{subfigure}[t]{0.48\textwidth}
    \centering
    \includegraphics[width=\textwidth]{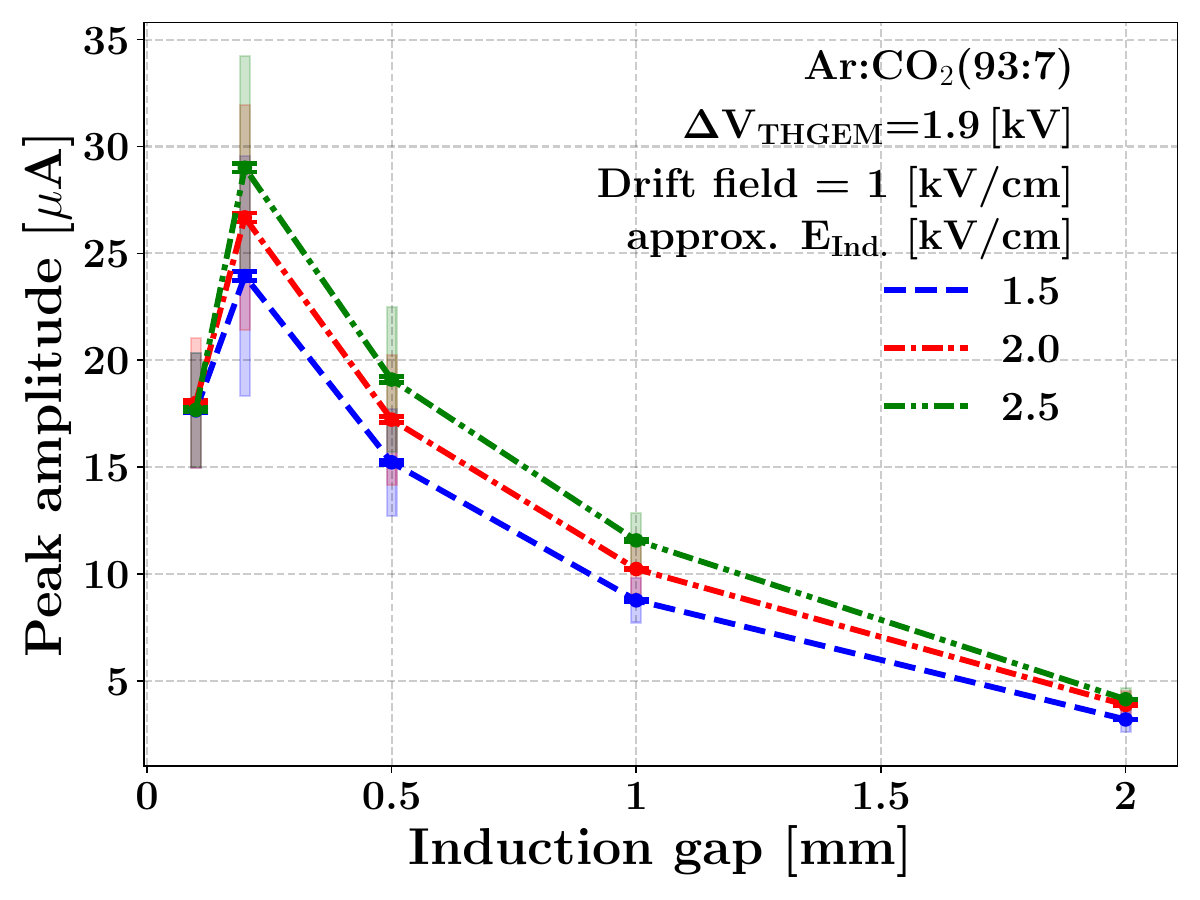}
    \caption{Peak amplitude.}
    \label{fig_peak_amplitude_induction_gap}
  \end{subfigure}
  \hfill
  \begin{subfigure}[t]{0.48\textwidth}
    \centering
    \includegraphics[width=\textwidth]{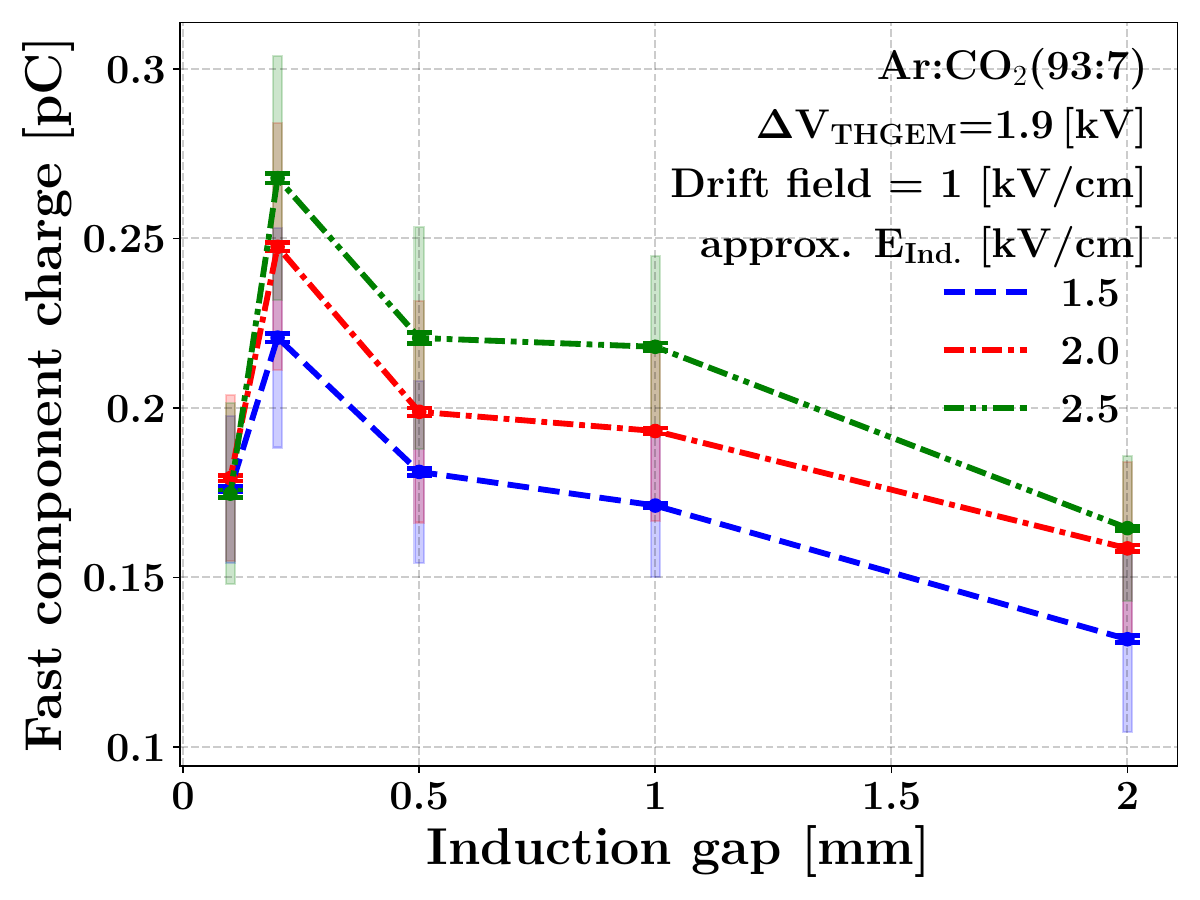}
    \caption{Charge of the fast component.}
    \label{fig_charge_fast_component_induction_gap}
  \end{subfigure}

  \caption{Current-signal properties for different induction gaps.}
  \label{fast_amplifier_induction_gap}
\end{figure}

The signal \textit{rise time}, \textit{pulse duration}, \textit{peak amplitude}, and \textit{fast-component charge} as functions of the induction-gap value and field, are shown in Figure~\ref{fast_amplifier_induction_gap}. Three induction field configurations were evaluated, corresponding to 1.5, 2, and 2.5~kV/cm under the parallel-plate approximation.
The fastest rise time, shortest pulse duration, largest peak amplitude and fast-component charge are observed for the 0.2~mm induction gap for all studied induction fields.
The consistent improvement in all waveform parameters indicates that an induction gap of 0.2~mm provides the optimal signal characteristics among the configurations investigated.

\paragraph{Charge signals}

\begin{figure}[htbp]
    \centering
    \begin{subfigure}[b]{0.48\textwidth}
    \includegraphics[width=\linewidth]{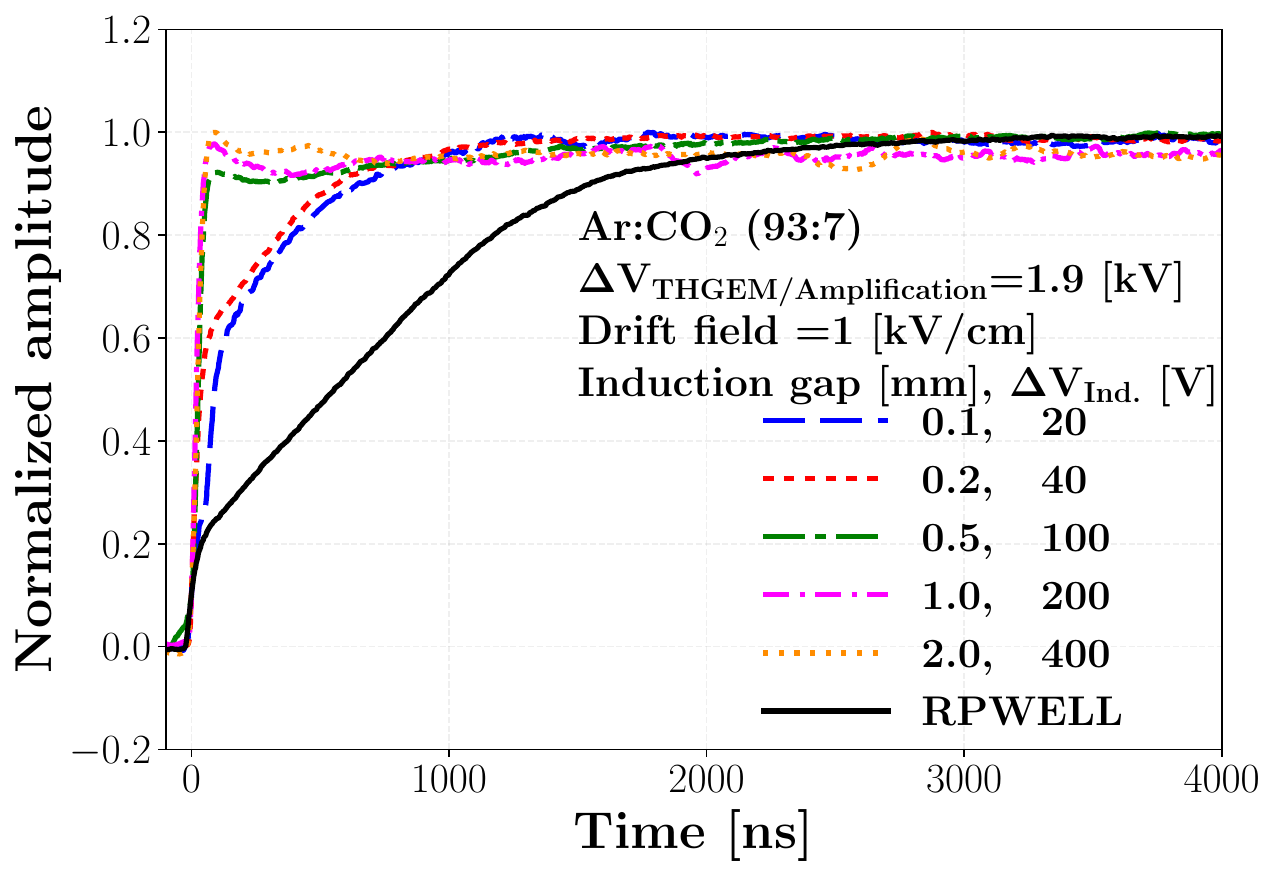}
    \caption{Normalized anode charge waveforms for RPWELL and THGEM with different induction-gap values.}
    \label{csp_all_gaps}
    \end{subfigure}
    \hfill
    \begin{subfigure}[b]{0.48\textwidth}
    \includegraphics[width=\linewidth]{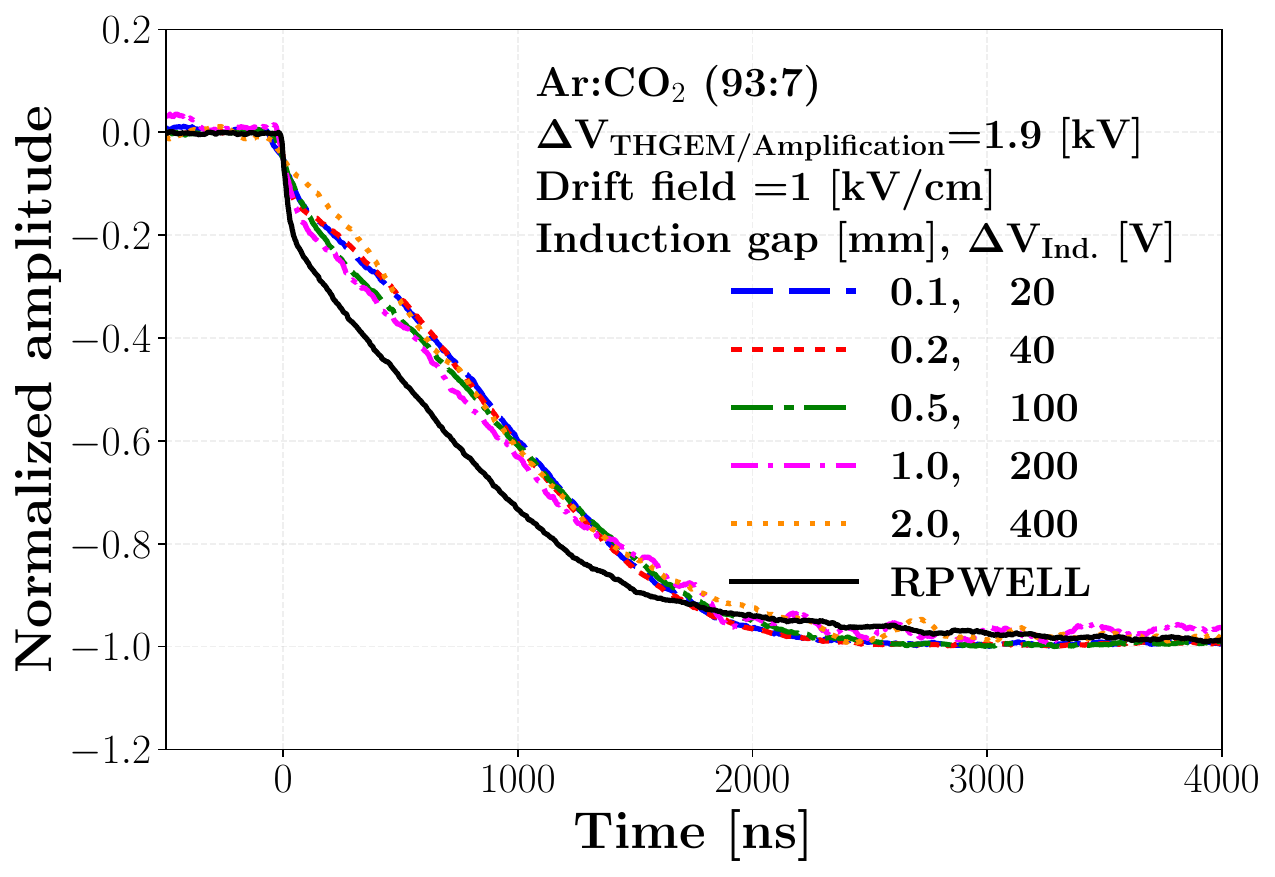}
    \caption{Normalized charge signal waveforms for RPWELL-top and THGEM-top different induction-gap values.}
    \label{csp_all_top_gaps}
    \end{subfigure}
    \caption{Typical charge waveforms.}
    \label{signal_waveforms_charge}
\end{figure}

Charge signals from the anode and THGEM-top electrodes are shown in Figures~\ref{csp_all_gaps} and~\ref{csp_all_top_gaps}, respectively. The THGEM-top waveform is almost unaffected by the gap size, and it is similar to the RPWELL one~\cite{rubin2013first}.
Anode charge signals are dominated by a slow ion component for induction-gap values smaller than 0.5~mm, while configurations with induction gaps larger than 0.5~mm are dominated by the faster electron component. In any case, the THGEM anode waveforms are significantly faster than RPWELL ones, due to the THGEM-bottom electrode screening ion movement above it.

\begin{figure}[htbp]
  \centering
    \begin{subfigure}[b]{0.48\textwidth}
    \centering
    \includegraphics[width=\textwidth]{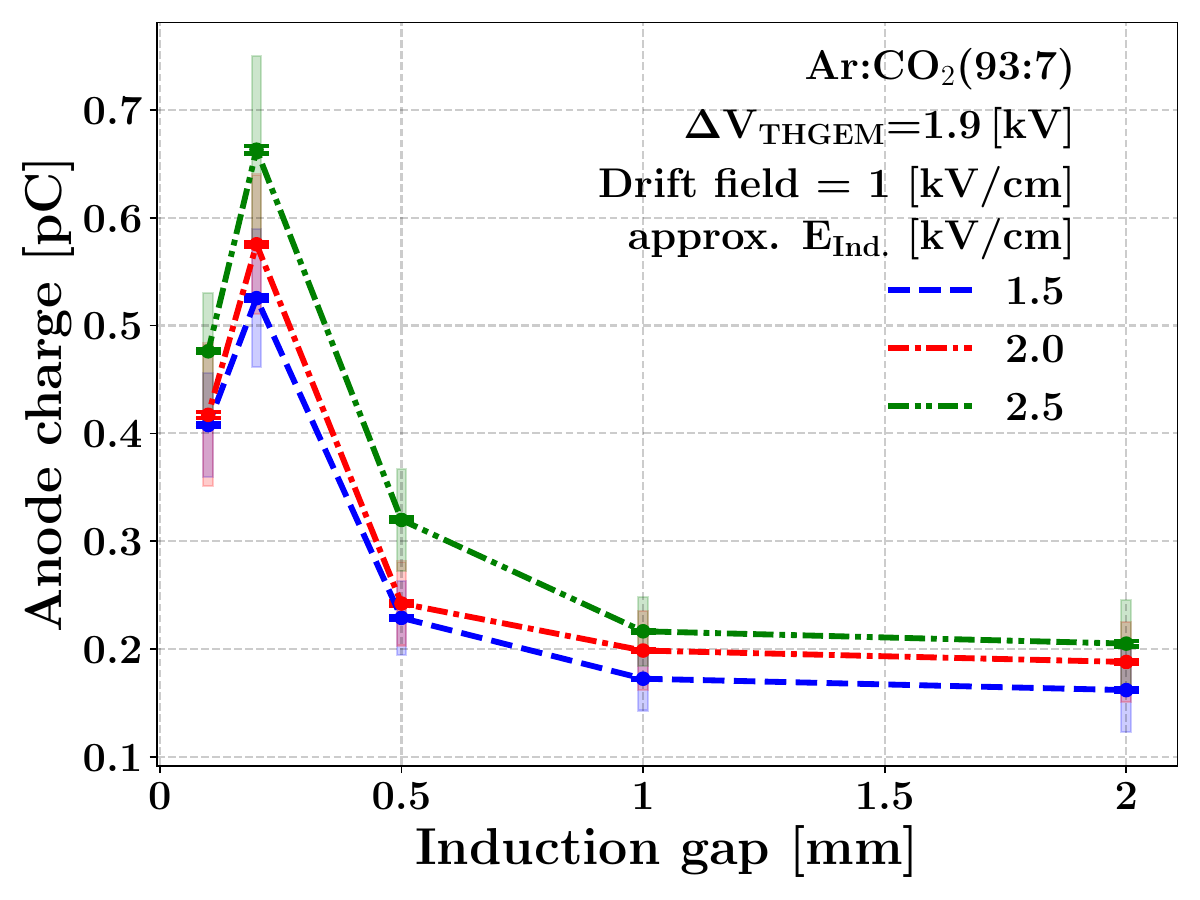}
    \caption{Anode charge.}
    \label{fig_anode_charge_induction_gap_csp}
  \end{subfigure}
  \hfill
    \begin{subfigure}[b]{0.48\textwidth}
    \centering
    \includegraphics[width=\textwidth]{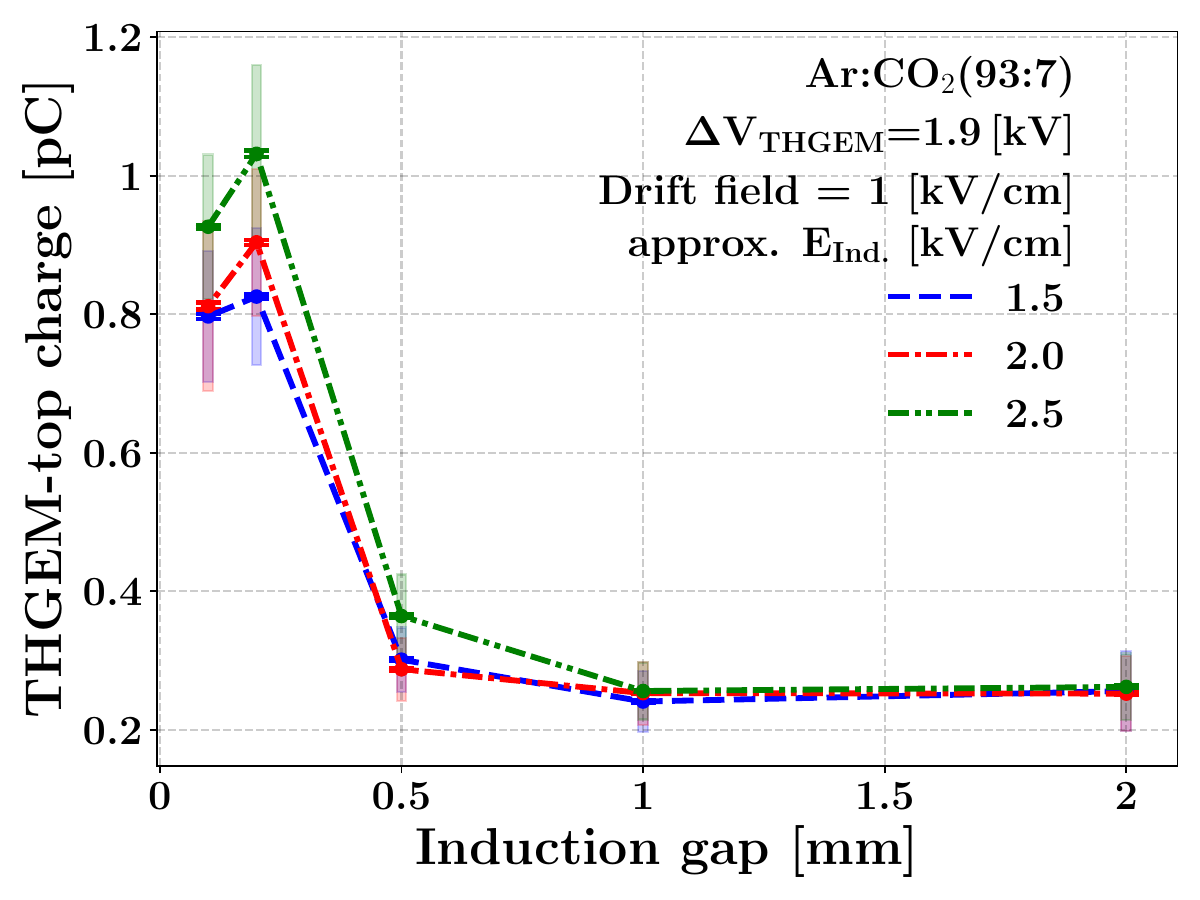}
    \caption{Top charge.}
    \label{fig_top_charge_induction_gap_csp}
  \end{subfigure}
  \vspace{0.5cm}
  \begin{subfigure}[b]{0.48\textwidth}
    \centering
    \includegraphics[width=\textwidth]{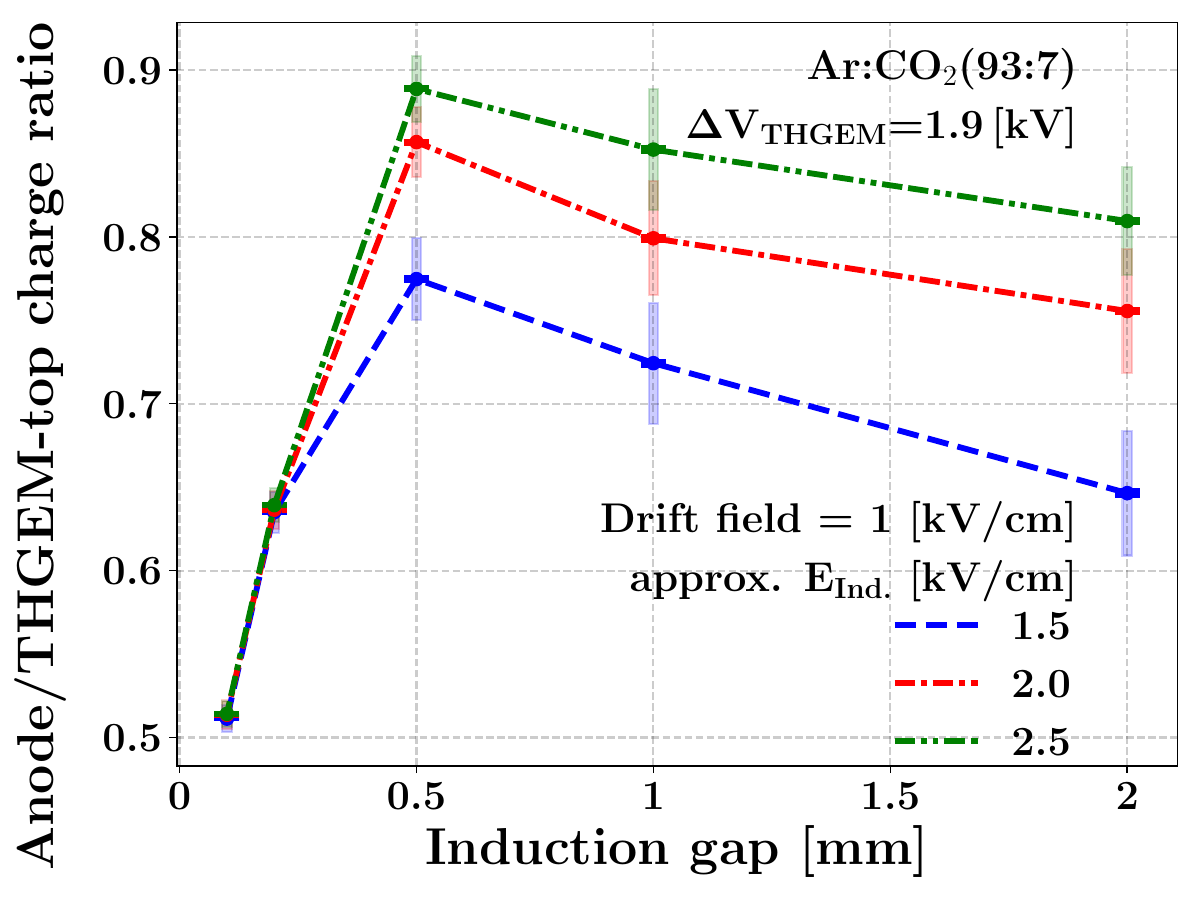}
    \caption{Charge ratio.}
    \label{fig_charge_ratio_induction_gap_csp}
  \end{subfigure}
  \hfill
  \begin{subfigure}[b]{0.48\textwidth}
    \centering
    \includegraphics[width=\textwidth]{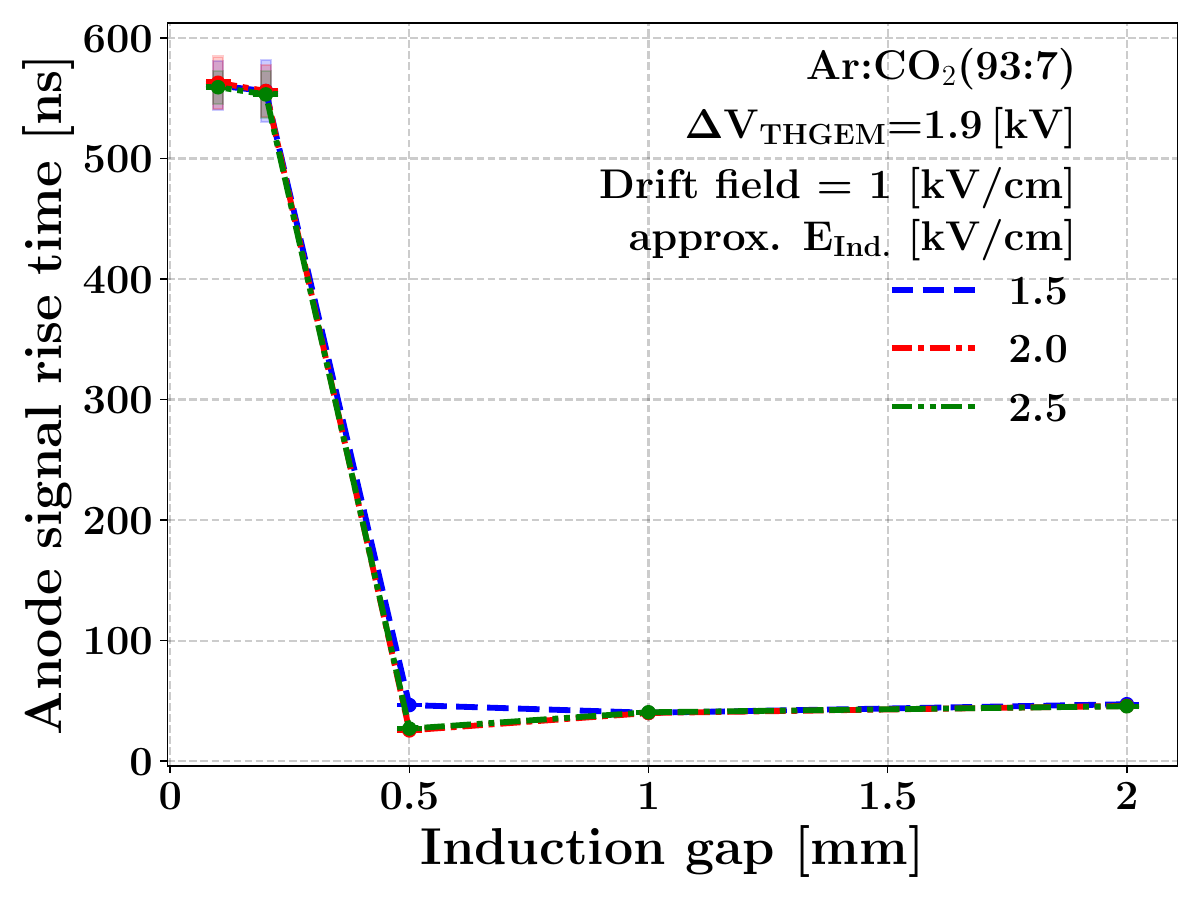}
    \caption{Rise time.}
    \label{fig_rise_time_induction_gap_csp}
  \end{subfigure}
  \hfill

  \caption{Charge-signal properties for different induction gaps.}
  \label{csp_induction_gap}
\end{figure}

The charge-signal properties as functions of different induction-gap values are shown in Figure~\ref{csp_induction_gap}. Figure~\ref{fig_anode_charge_induction_gap_csp} shows that the maximum anode charge is obtained with a 0.2~mm induction gap. A similar trend is observed for the THGEM-top charge, as shown in Figure~\ref{fig_top_charge_induction_gap_csp}.

The charge ratio of the anode charge to the THGEM-top charge, which we interpret as an approximation of the charge reaching the anode with respect to the total avalanche charge, is shown in Figure~\ref{fig_charge_ratio_induction_gap_csp}. It is highest for the 0.5~mm induction gap, relatively flat for larger gap values, and decreases to about 0.65 for the 0.2~mm gap. This loss is compensated by the overall larger avalanche gain for this gap value.
The different field configurations tested show that a higher field in the induction region causes both a higher avalanche gain and a larger fraction of electrons reaching the anode.
Figure~\ref{fig_rise_time_induction_gap_csp} shows a sharp transition from a long rise time to a short rise time between the 0.2~mm and 0.5~mm induction gaps. This is attributed to signal induction being dominated by ion motion for the 0.2~mm gap and by electron motion for the 0.5~mm gap.

\subsection{Study with 0.2~mm induction gap and different field configurations}

Following the results in section~\ref{sec:Study_of_induction_gap_and_field_configurations}, an induction gap of 0.2~mm was observed to be optimal for both fast current signals and charge gain maximization. This configuration was therefore studied with different \DV{THGEM} and \DV{induction} values.

\begin{figure}[htbp]
\centering
        \includegraphics[width=\linewidth]{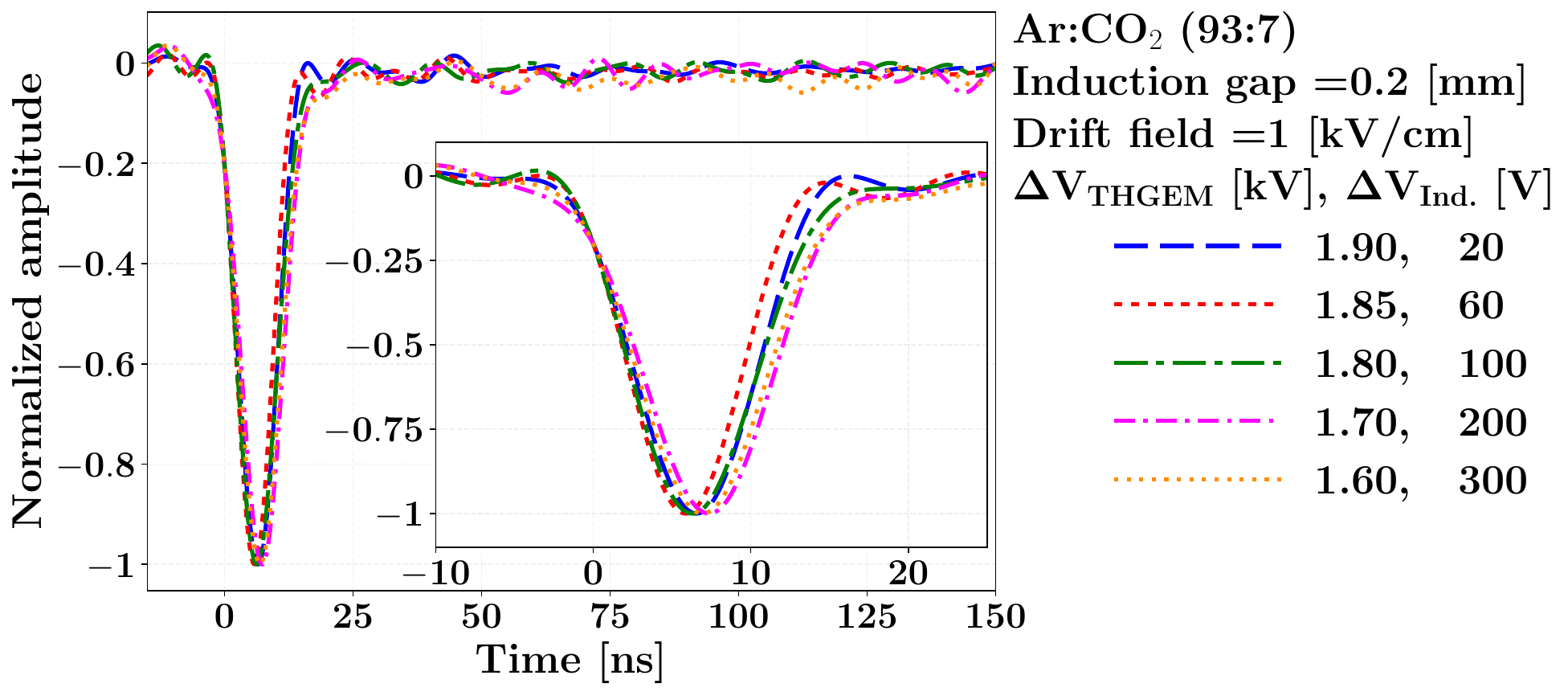}
        \caption{Typical THGEM current signal with a 0.2~mm induction gap for different \DV{THGEM} and \DV{induction} values.}
        \label{fca_scan_THGEM_current_all_induction}
\end{figure}

\begin{figure}[htbp]
  \centering
  \begin{subfigure}[b]{0.48\textwidth}
    \centering
    \includegraphics[width=\textwidth]{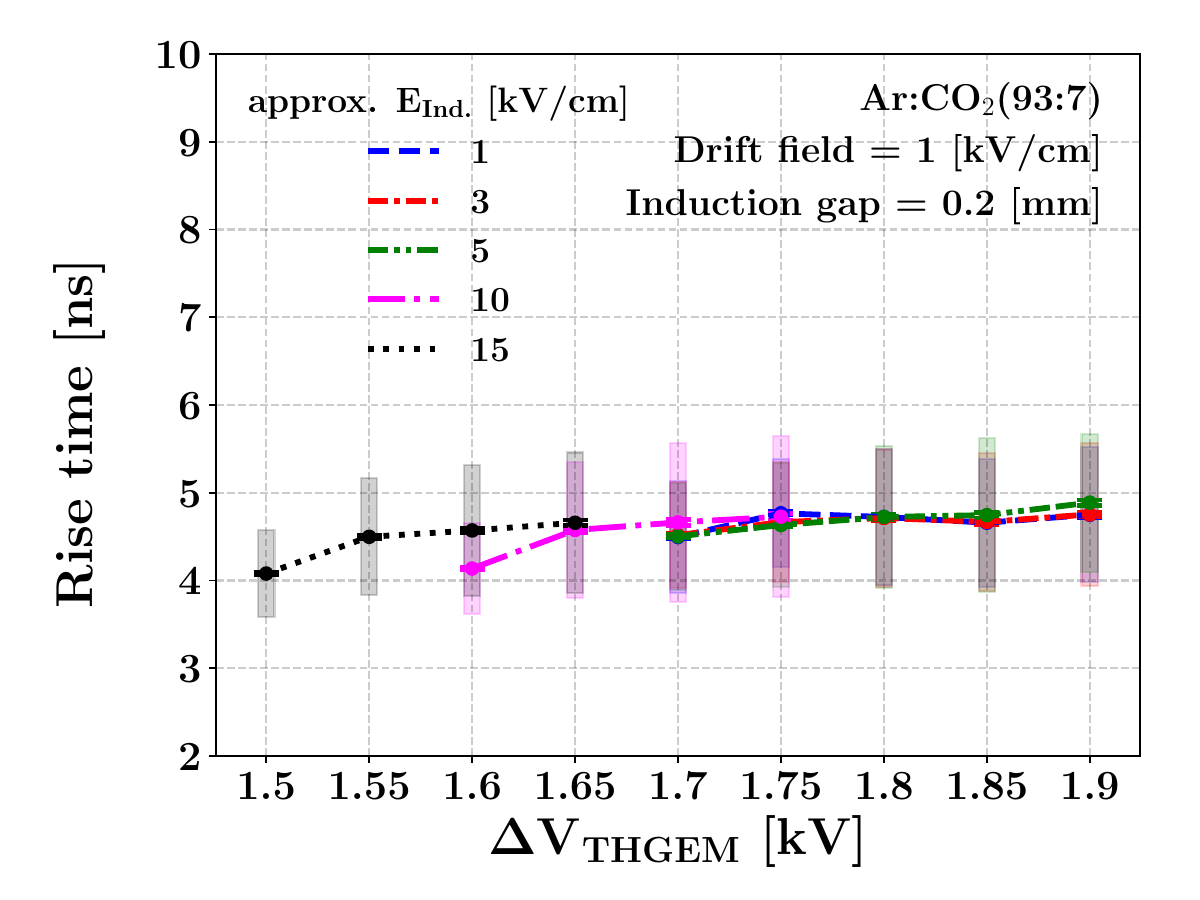}
    \caption{Rise time.}
    \label{fig_rise_time_current_scan}
  \end{subfigure}
  \hfill
  \begin{subfigure}[b]{0.48\textwidth}
    \centering
    \includegraphics[width=\textwidth]{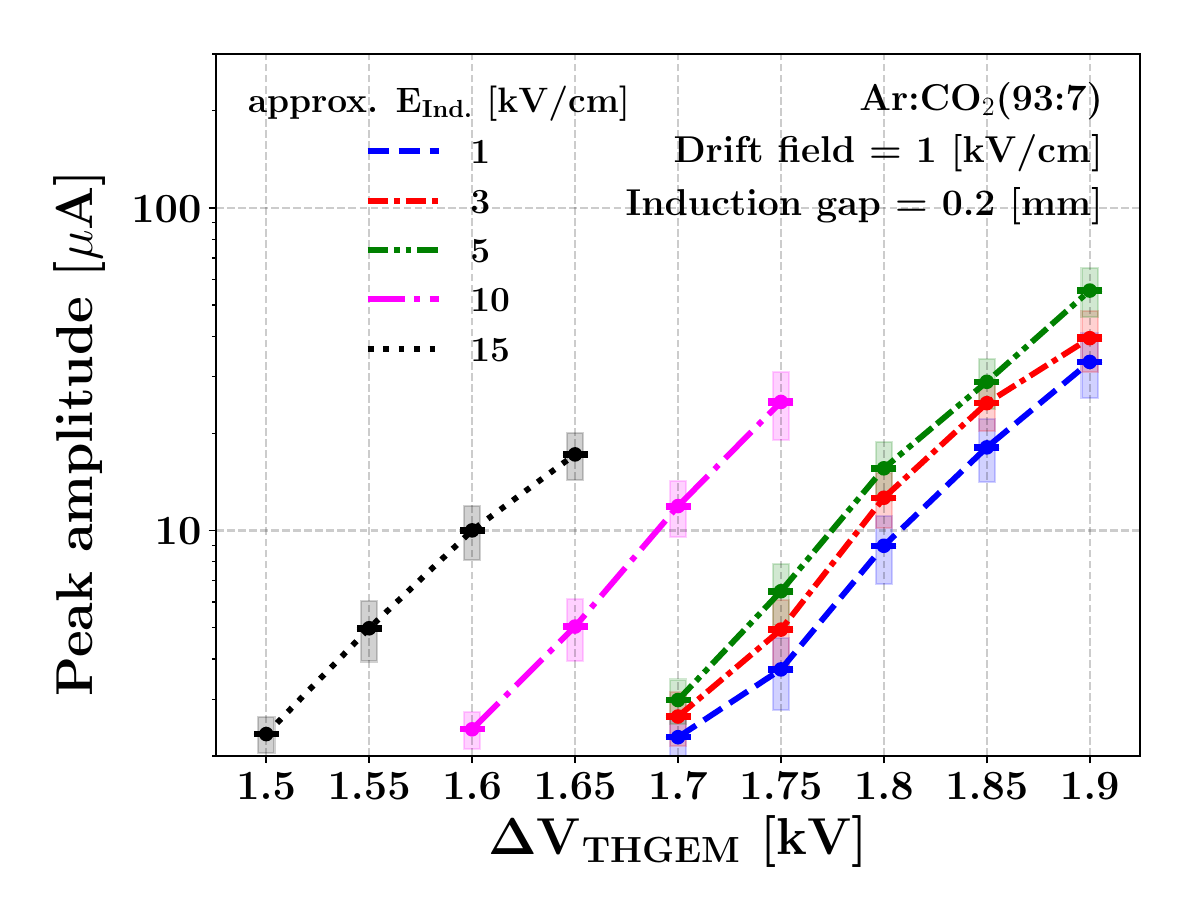}
    \caption{Peak amplitude.}
    \label{fig_peak_amplitude_current_scan}
  \end{subfigure}
  \begin{subfigure}[b]{0.48\textwidth}
    \centering
    \includegraphics[width=\textwidth]{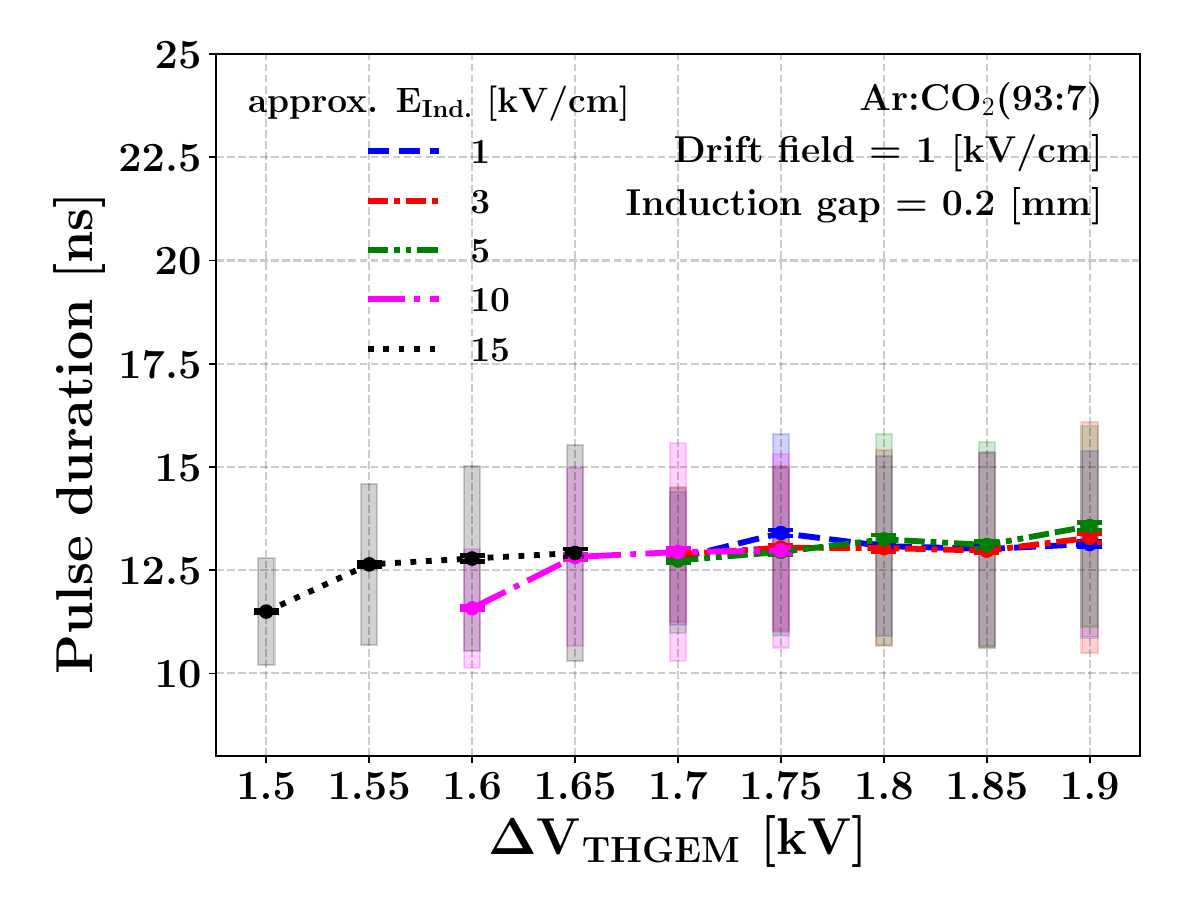}
    \caption{Pulse duration.}
    \label{fig_time_interval_current_scan}
  \end{subfigure}
  \hfill
  \begin{subfigure}[b]{0.48\textwidth}
    \centering
    \includegraphics[width=\textwidth]{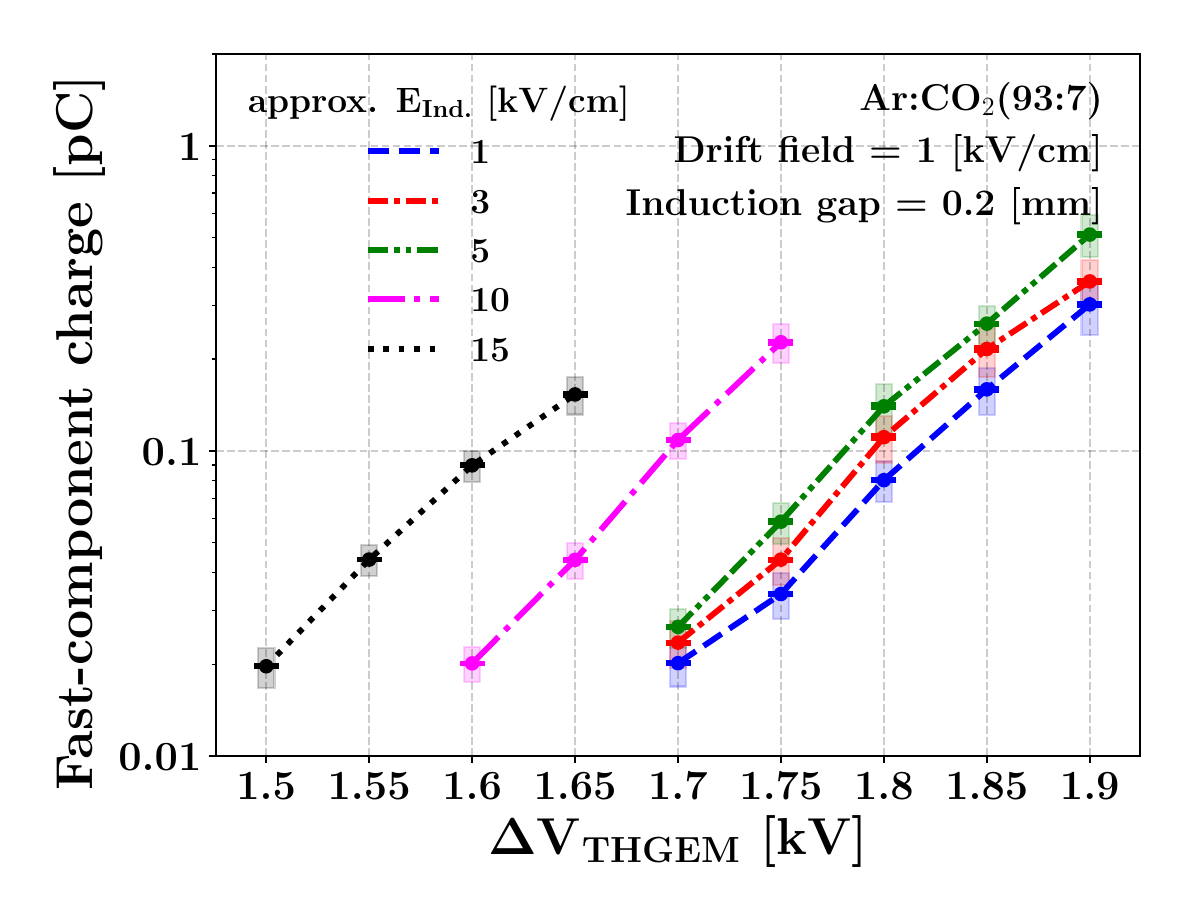}
    \caption{Fast-component charge.}
    \label{fig_fast_component_charge_current_scan}
  \end{subfigure}

  \caption{Current-signal properties with a 0.2~mm induction gap for different field configurations.}
  \label{current_field_scan}
\end{figure}

\paragraph{Current signals}
Typical current signal waveforms for different \DV{THGEM} and \DV{induction} values to obtain a similar avalanche gain of $\sim$1.3$\times10^{4}$ are shown in Figure~\ref{fca_scan_THGEM_current_all_induction}.
The results for the \textit{rise time}, \textit{peak amplitude}, \textit{pulse duration}, and \textit{fast-component charge} at different \DV{THGEM} and \DV{induction} are shown in Figure~\ref{current_field_scan}.
As seen in Figure~\ref{electric_field_200_um}, the electric fields up to the center of the hole from the cathode, where they reach their maximum values, are nearly identical for all induction fields. Therefore, the observed differences are mainly associated with variations in the field configuration from the center of the hole to the anode. This has two consequences. First, the rise time (Figure~\ref{fig_rise_time_current_scan}) and pulse duration (Figure~\ref{fig_time_interval_current_scan}), which are governed by electron motion and begin once the number of electrons is sufficient to induce a signal above the noise level, are only slightly affected by the induction field. Second, since electrons produced in the induction gap are rapidly collected at the anode and contribute only marginally to the induced signal, the peak amplitude (Figure~\ref{fig_peak_amplitude_current_scan}) and  fast-component charge (Figure~\ref{fig_fast_component_charge_current_scan})  are very similar for induction fields of 1--5~kV/cm, where the fields inside the hole are nearly identical. In contrast, the larger differences in the field inside the hole for induction fields of 10--15~kV/cm enhance charge production within the hole, thereby affecting both the peak amplitude and the fast-component charge.

\begin{figure}[htbp]
     \begin{subfigure}[b]{0.48\textwidth}
    \includegraphics[width=\linewidth]{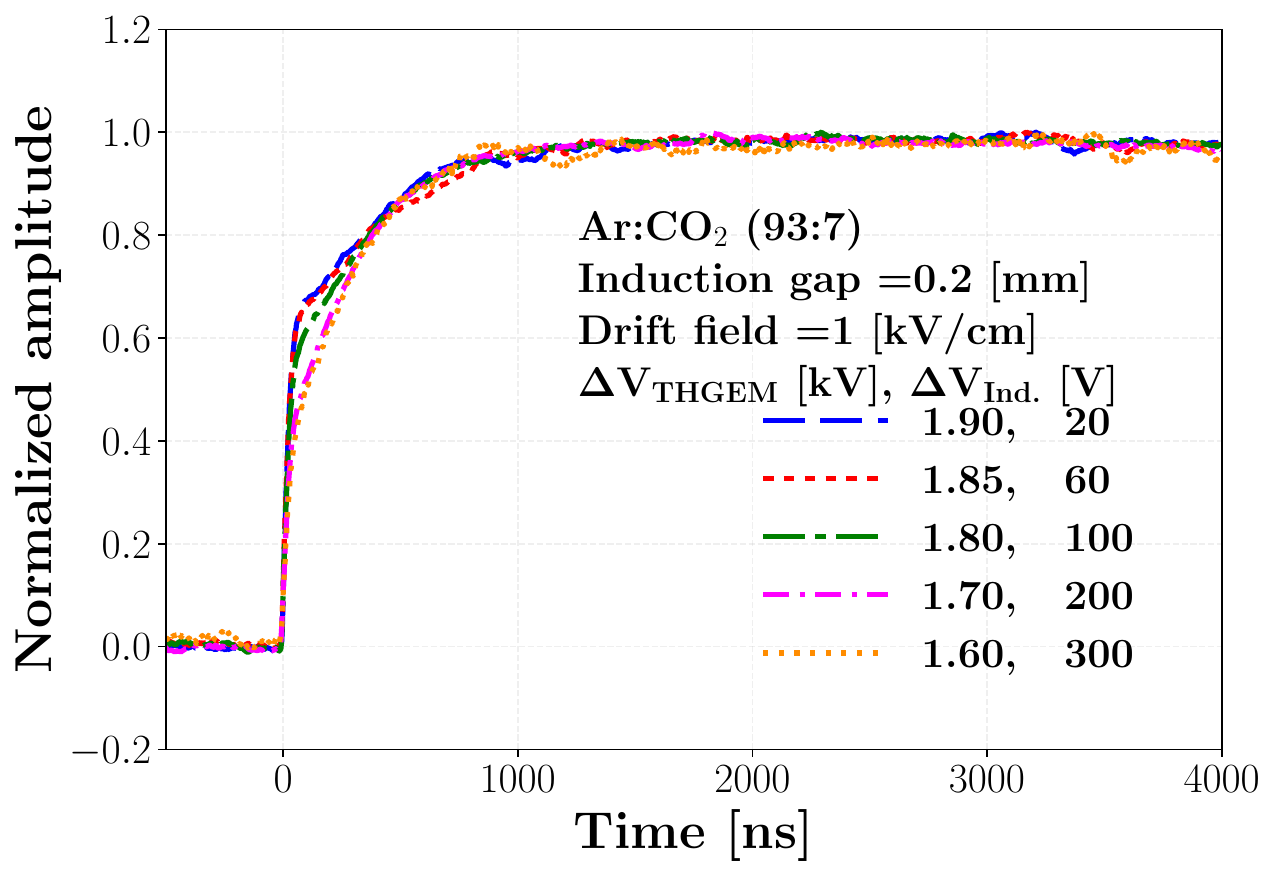}
    \caption{Normalized anode charge waveforms for a 0.2~mm induction gap.}
    \label{csp_anode_200um}
    \end{subfigure}
    \hfill
    \begin{subfigure}[b]{0.48\textwidth}
    \includegraphics[width=\linewidth]{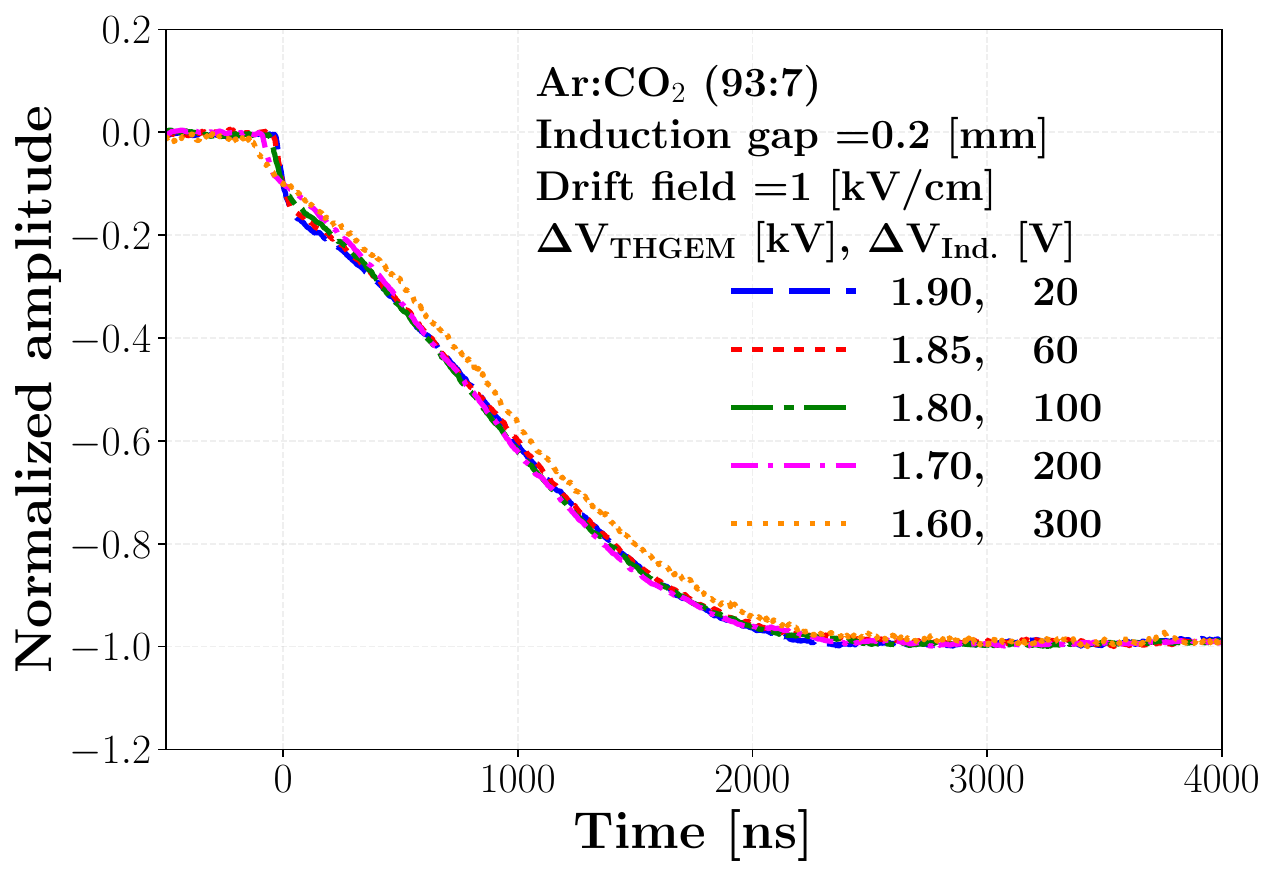}
    \caption{Normalized THGEM-top charge waveforms for a 0.2~mm induction gap.}
    \label{csp_top_200um}
    \end{subfigure}
\caption{Charge signals for the 0.2~mm induction gap configuration}
\label{charge_Sginal_field_scan}
\end{figure}

\begin{figure}[htbp]
  \centering
  \begin{subfigure}[b]{0.48\textwidth}
    \centering
    \includegraphics[width=\textwidth]{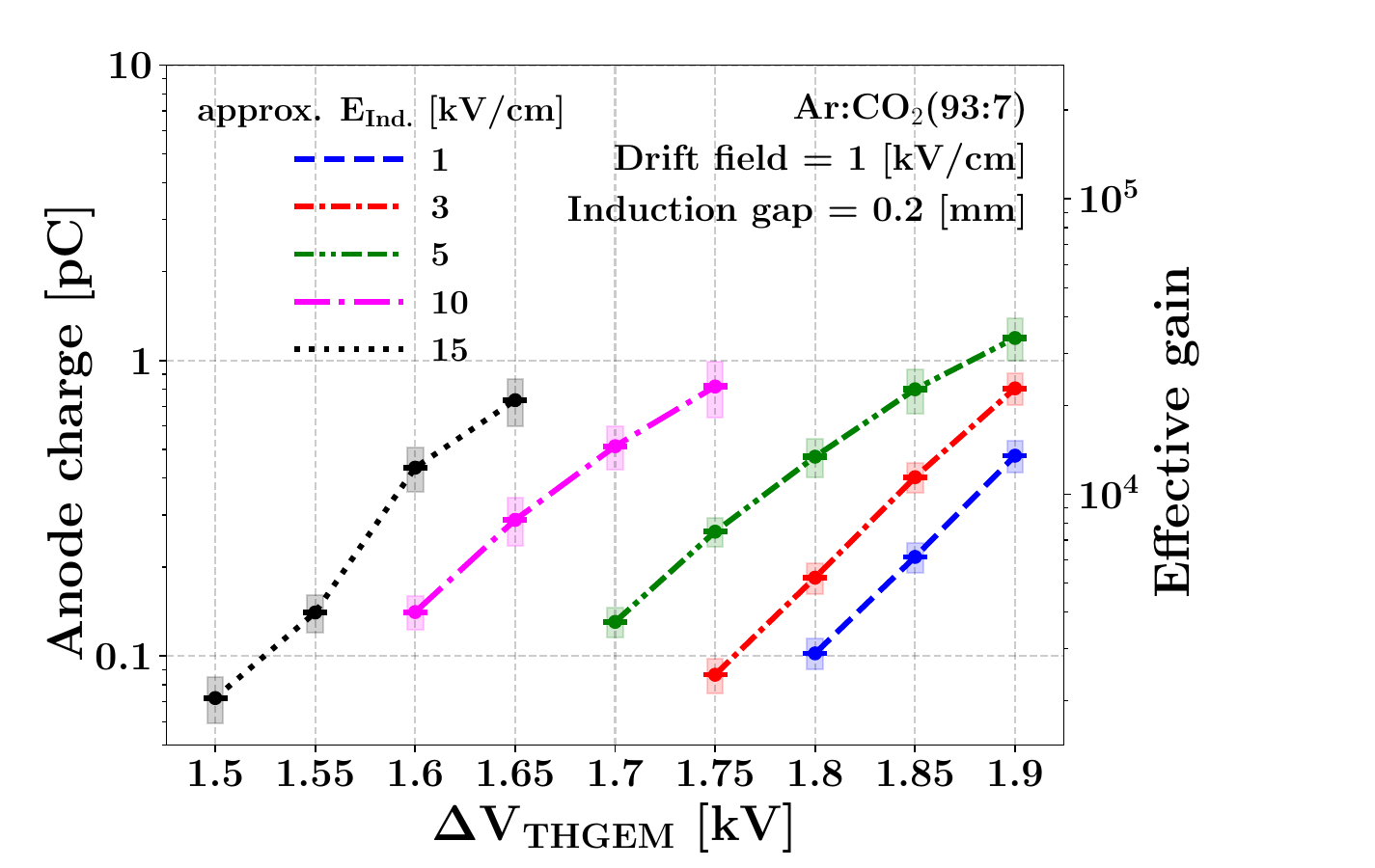}
    \caption{Anode charge.}
    \label{fig_anode_charge_scan}
  \end{subfigure}
  \hfill
  \begin{subfigure}[b]{0.48\textwidth}
    \centering
    \includegraphics[width=\textwidth]{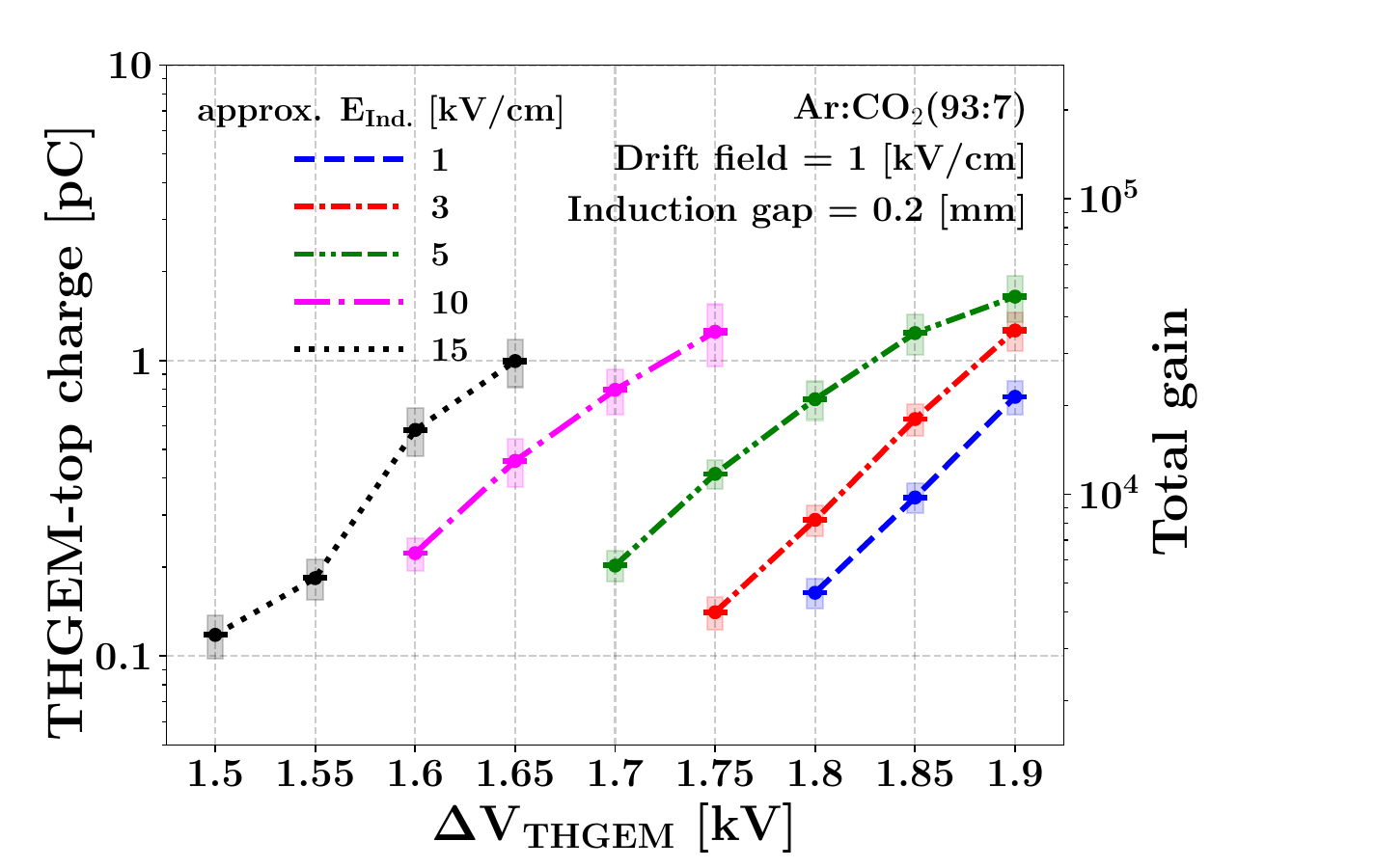}
    \caption{THGEM-top charge.}
    \label{fig_top_charge_scan}
  \end{subfigure}
  \vspace{0.5cm}
 \begin{subfigure}[b]{0.48\textwidth}
    \centering
    \includegraphics[width=\textwidth]{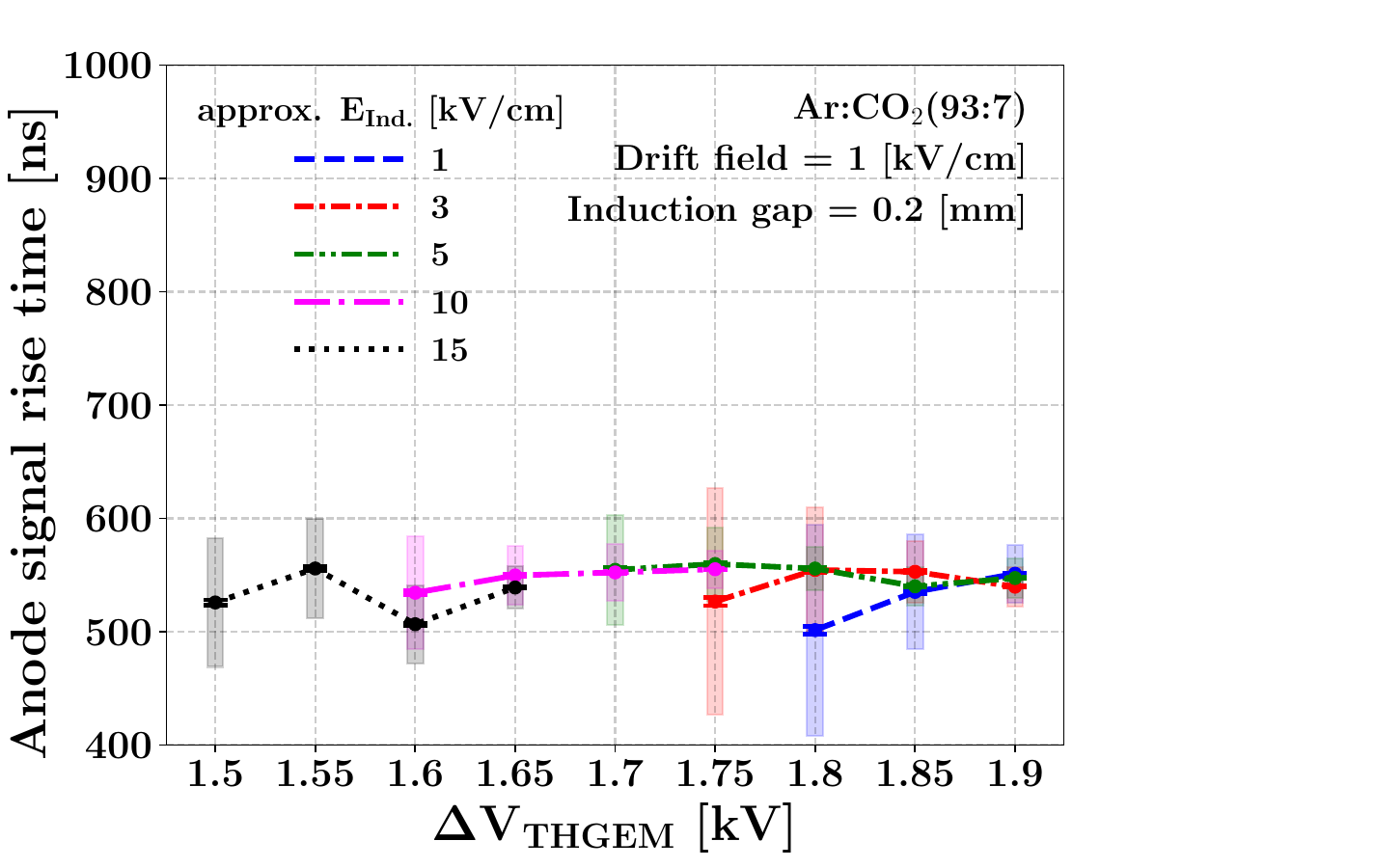}
    \caption{Rise time.}
    \label{fig_rise_time_charge_scan}
  \end{subfigure}
  \hfill
  \begin{subfigure}[b]{0.48\textwidth}
    \centering
    \includegraphics[width=\textwidth]{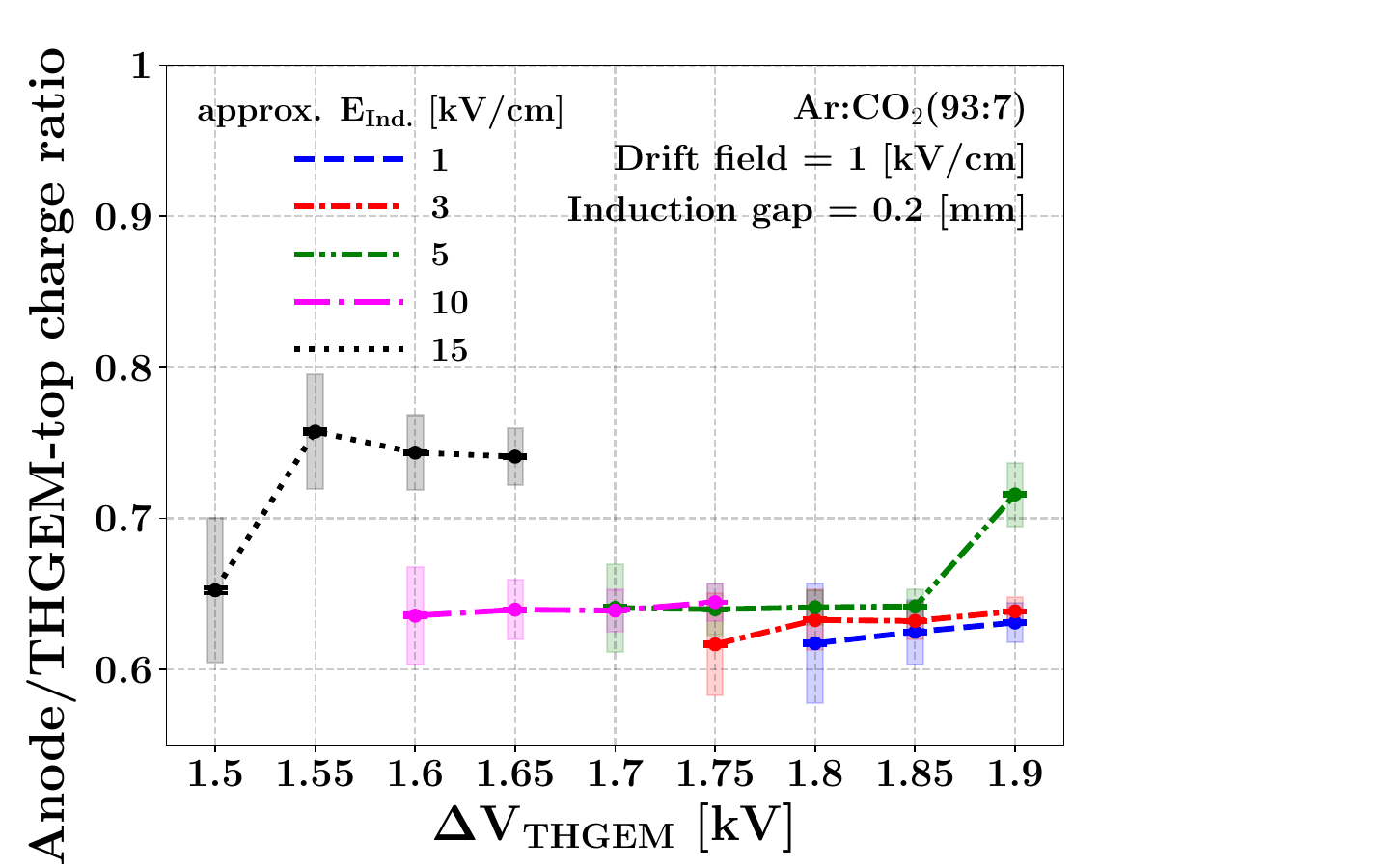}
    \caption{Anode/THGEM-top charge ratio.}
    \label{fig_charge_ratio_scan}
  \end{subfigure}
  \caption{Charge-signal properties with a 0.2~mm induction gap for different field configurations.}
  \label{charge_field_scan}
\end{figure}

\paragraph{Charge signals}
Typical charge-signal waveforms for different \DV{THGEM} and \DV{induction} values to obtain a similar avalanche gain of $\sim$1.3$\times10^{4}$ are shown for the anode and THGEM top electrode in Figure~\ref{csp_anode_200um} and Figure~\ref{csp_top_200um}, respectively.
The corresponding results for \textit{anode charge}, \textit{THGEM-top charge}, \textit{anode/THGEM-top charge ratio} and \textit{anode signal rise time} are shown in Figure~\ref{charge_field_scan}. The rise time of the charge waveform is similar for different field configurations, and the charge ratio is also similar for different induction fields and \DV{THGEM} values, except for 15~kV/cm, in the parallel-plate-equivalent approximation, where a higher ratio of anode charge to top charge is observed, indicating better charge collection at the anode due to the high induction field.

\subsection{Time resolution with cosmic rays}
\label{time_resolution}

\begin{figure}
    \centering
    \begin{subfigure}[b]{0.48\textwidth}
    \includegraphics[width=\linewidth]{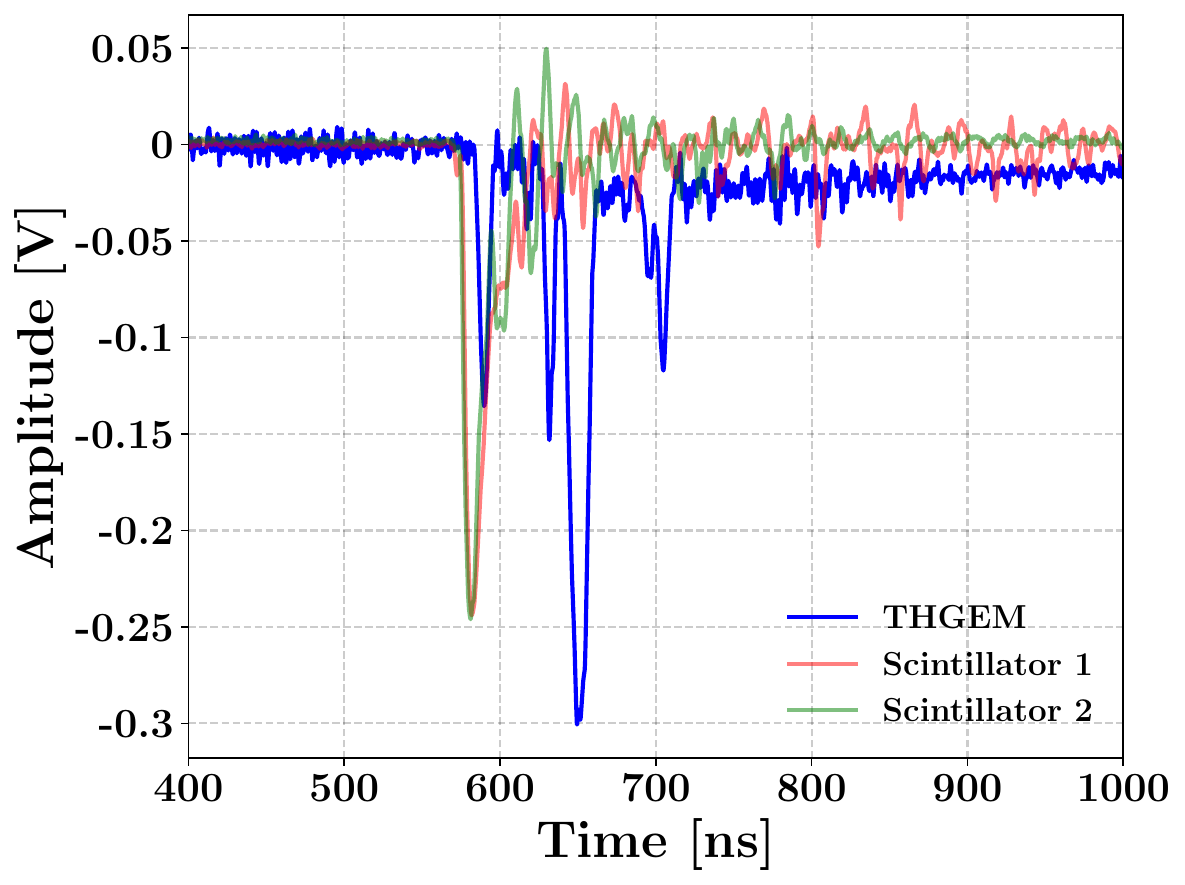}
    \caption{Typical current signals of the scintillators and the THGEM with cosmic muon with a 0.2~mm induction gap at \DV{THGEM} = 1900~V, \DV{induction} = 100~V, and a drift field of 1~kV/cm.}
    \label{thgem_clusters_with_cosmic_muon}
    \end{subfigure}
    \hfill
    \begin{subfigure}[b]{0.48\textwidth}
    \includegraphics[width=\linewidth]{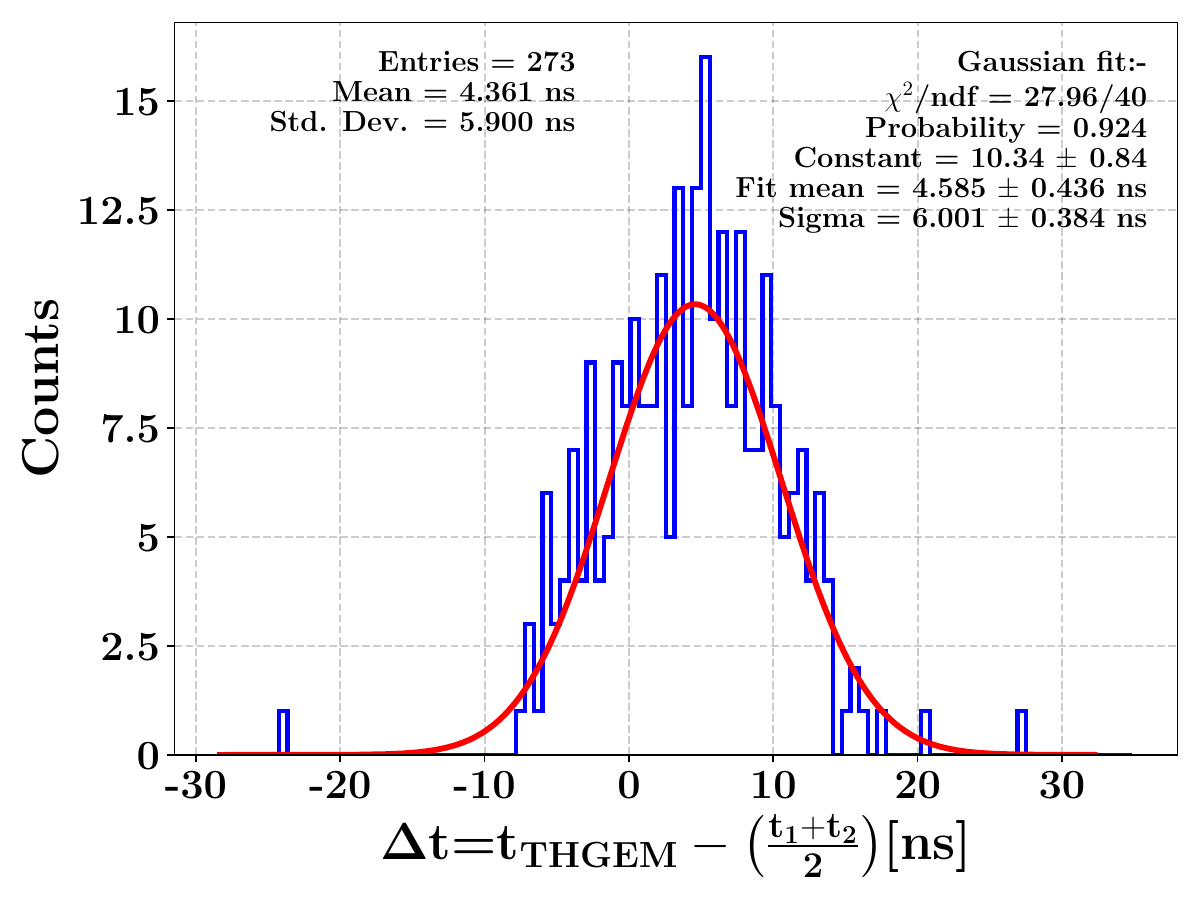}
    \caption{Time jitter distribution of cosmic muon current signals with a 0.2~mm induction gap at \DV{THGEM} = 1900~V, \DV{induction} = 100~V, and a drift field of 1~kV/cm.}
    \label{time_resolution_fit}
    \end{subfigure}
    \caption{Time-resolution measurement using cosmic muons for a THGEM with a 0.2~mm induction gap.}
\label{fig:time_resolution_cosmic_muons}
\end{figure}

This measurement was taken with 0.2~mm induction gap while the THGEM was operated at \DV{THGEM}=1900~V and \DV{induction}=100~V (approx. induction field of 5~kV/cm in parallel plate), resulting in the highest observed peak amplitude and fastest rise time.
The time-jitter distribution of the first peak in the THGEM current signal from cosmic muons is shown in Figure~\ref{time_resolution_fit}. The measured time resolution is 6.0~ns (standard deviation), dominated by variations of the distance between the closest primary ionization cluster and the THGEM-top electrode.

\subsection{Effect of resistive plate on electric discharges}
\label{discharge_quenching}

\begin{figure}[htbp]
    \centering
    \begin{subfigure}[t]{0.48\textwidth}
    \includegraphics[width=\linewidth]{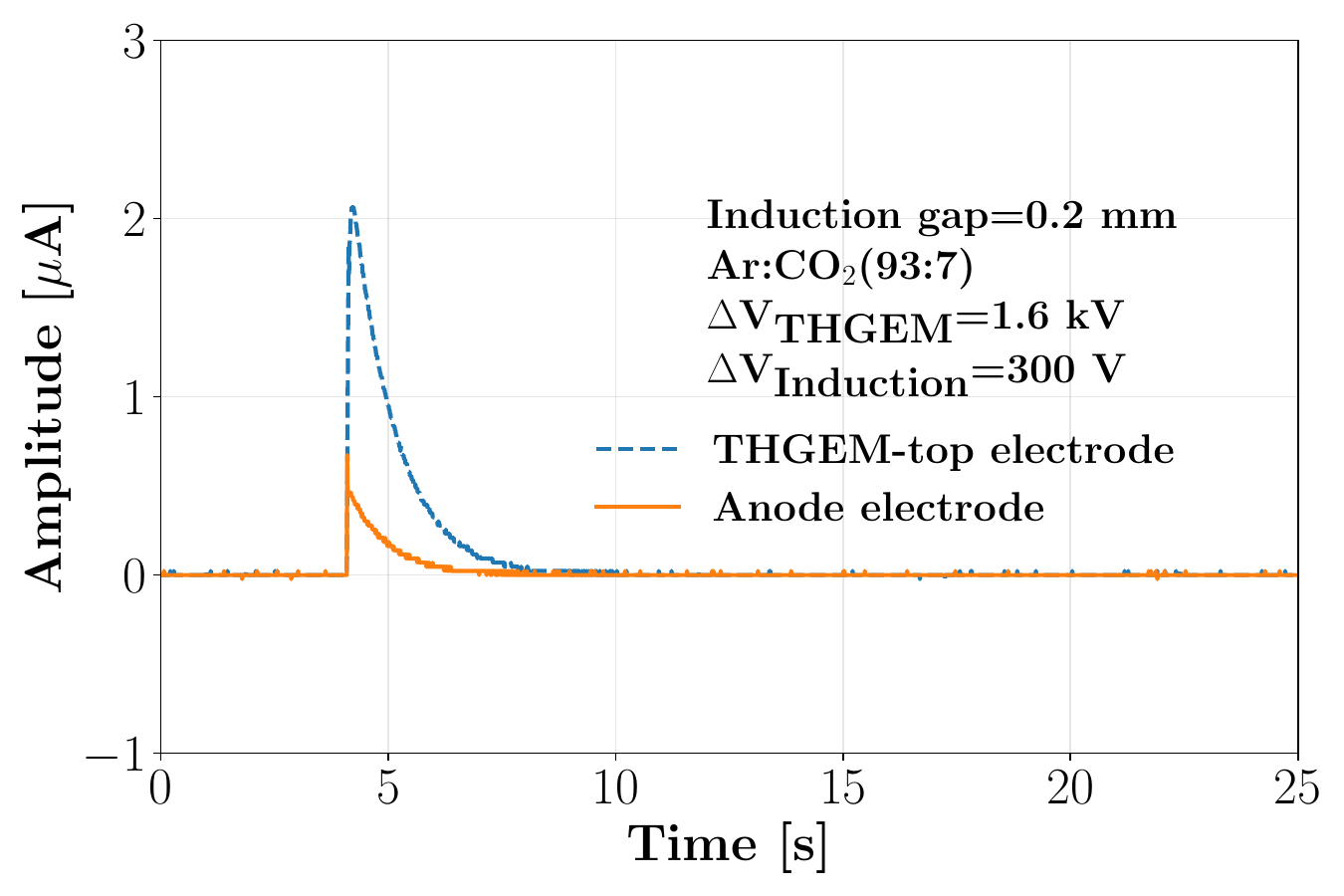}
    \caption{Discharge without the resistive plate.}    
    \end{subfigure}
    \hfill
    \begin{subfigure}[t]{0.48\textwidth}
    \includegraphics[width=\linewidth]{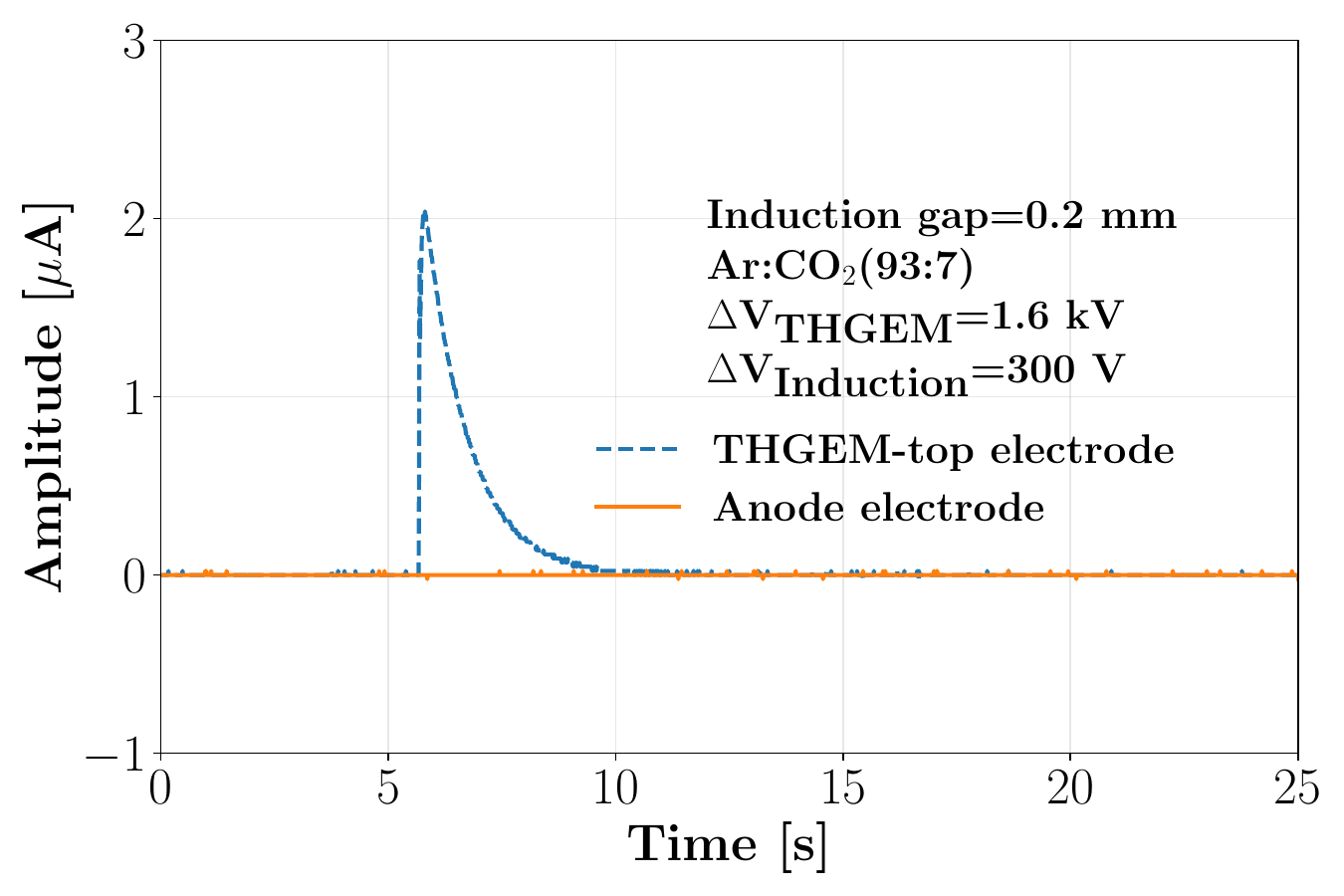}
    \caption{Discharge with the resistive plate.}    
    \end{subfigure}
    \caption{Power supply currents during a typical discharge with a 0.2~mm induction gap under irradiation with a \textsuperscript{55}Fe X-ray source. The voltage configuration corresponds to an induction field of 15~kV/cm (parallel plate approximation) and \DV{THGEM} = 1.6~kV.}
    \label{fig_discharge}
\end{figure}

As the induction gap is reduced, the resistive plate anode plays an increasingly important role in mitigating the effect of electrical discharges, enabling stable operation with very small induction gaps. Typical current monitor waveforms provided by the power supply during a discharge, recorded without and with a resistive plate coupled to the anode, are shown in Figure~\ref{fig_discharge}. It shows that when a resistive plate is coupled to the anode, the current that flows through it during a discharge is strongly suppressed.
A quantitative comparison of the discharge behavior is presented in Table~\ref{table_1} for different values of \DV{induction} and \DV{THGEM}, with and without the resistive plate, for a 0.2~mm induction gap. The discharge rate was determined in both configurations by monitoring discharges at the top electrode and measuring the pulse durations between ten discharge intervals. The rate is not affected by the presence of the resistive plate. The intensity of the discharge at each electrode was evaluated by integrating the current waveform. In the bare-anode configuration, the integrated current increases with both \DV{induction} and \DV{THGEM}, while the resistive plate effectively suppresses it in all configurations.

\begin{table}[htbp]
\centering
\caption{Discharge rate and current integral (charge) at the anode due to a discharge. Comparison between metal and resistive plate (RP) anodes}
\smallskip
\begin{tabular}{lccccc}
\hline
\hline
Configuration & E\textsubscript{ind.}~[kV/cm], \DV{THGEM}~[V] & \multicolumn{2}{c}{Discharge rate~[Hz]} & \multicolumn{2}{c}{Charge at anode~[$\mu$C]}\\
& 
& Metal & RP
& Metal & RP\\
\hline
1 & 1  ,  1900& 0.00134 &0.00092& 0.04062&-\\
2 & 3  ,  1850& 0.00069&0.00075& 0.09267& 0.01425\\
3 & 5  ,  1800& 0.00145&0.00080& 0.12343&0.01139\\
4 &10  ,  1700& 0.00032&0.00238& 0.27473&-\\
5 &15  ,  1600& 0.00022&0.00492& 0.42048&-\\
\hline
\hline
\end{tabular}
\label{table_1}
\end{table}

\section{Discussion}
\label{sec:discussion}

In this work, a THGEM detector coupled to a resistive anode was characterized for the first time.  This allows stable operation with small induction gap in a high field regime that would otherwise be strongly limited by discharge effects in conventional metallic anode configurations, enabling the simultaneous realization of fast signals and significant charge amplification. 

Previous studies of THGEM detectors with metallic anodes were typically limited to induction gaps of several hundred micrometers, generally in the range of 0.4--0.8~mm. By contrast, the resistive anode enabled stable operation with induction gaps down to $\sim$100~$\upmu\mathrm{m}$, opening a previously unexplored operating regime and allowing a systematic investigation of the induction-gap dependence of signal formation. This made it possible to identify operating conditions that provide enhanced signal amplitude, shorter rise times, and improved timing performance. 

For smaller gaps, the electron drift distance between the THGEM and the anode is shortened, resulting in faster electron collection and consequently improved timing characteristics.
The rise time of the THGEM current waveforms improves significantly, from $\sim$13~ns to $\sim$5~ns, as well as the current pulse amplitude, as the induction gap is reduced. In this configuration the current pulse is mainly due to electron movement within the THGEM hole.

In addition, reducing the induction gap leads to an increase in the electric-field strength in the induction region, which can promote additional charge multiplication at sufficiently high fields. Extending in this way the amplification region beyond the THGEM holes and into the induction gap results in an increased maximum gain.
However, when amplification occurs closer to the anode, a more pronounced ion contribution to the anode charge signal is observed. This manifests as a longer ion tail and a slower rise time in the CSP signals. This effect must be considered when optimizing detector operation for a specific integration time.

Overall, this study shows that combining small induction gaps with a resistive anode provides a practical way to tune both the charge and timing characteristics of THGEM-based detectors. The 0.2~mm induction gap represents the best compromise among the configurations studied, providing fast current signals, enhanced peak amplitude, and stable operation in a high-field induction region.

A next step for this concept could include a THGEM electrode with a resistive layer on the top side, similar to the one reported in~\cite{Song_2020}, instead of a metallic electrode. Such a configuration could allow effective discharge quenching also for larger detector areas, while maintaining efficient signal induction between the conductive bottom electrode and the resistive anode.

Stable operation at high gain and fast signal formation resulting from the combination of a small induction gap and a resistive-plate anode make this concept suitable for tracking and particle-counting applications, such as muon spectrometers or sampling elements in digital hadronic calorimetry~\cite{moleri2017resistive,zavazieva2023towards}. The capability to distinguish between current spikes produced by different ionization clusters could be used for particle identification by cluster-counting techniques.

\acknowledgments

The authors would like to thank Dr. Eraldo Oliveri and Dr. Djunes Janssens for their insightful comments and constructive suggestions. This work was supported by the Pazy Foundation, the  Krenter Perinot Center for High-Energy Particle Physics, the Israel Science Foundation(ISF), and the Nella and Leon Benoziyo Center for High Energy Physics. The authors also wish to express their sincere gratitude to Martin Kushner Schnur for his invaluable support of this research.

\bibliographystyle{elsarticle-num}                 
\bibliography{bibliography.bib}

\end{document}